\documentclass[prl,10pt,notitlepage,aps,twocolumn,superscriptaddress,nobibnotes,longbibliography]{revtex4-1}
\usepackage{amsmath}
\usepackage{amssymb}
\usepackage{bm}
\usepackage{dcolumn}
\usepackage{graphicx}
\usepackage{microtype}
\usepackage{multirow}
\usepackage[labeled,resetlabels]{multibib}
\newcites{S}{Supplementary material}
\bibliographystyleS{apsrev4-1}
\usepackage{hyperref}

\newcommand{\bk}{{\bm{k}}}
\newcommand{\up}{\mathord{\uparrow}}
\newcommand{\dn}{\mathord{\downarrow}}
\newcommand*\abs[1]{\lvert#1\rvert}
\DeclareMathOperator\real{Re}

\renewcommand{\arraystretch}{1.5}

\begin{document}

\title{Stabilizing even-parity chiral superconductivity in Sr$_2$RuO$_4$}

\author{Han Gyeol Suh}
\affiliation{Department of Physics, University of Wisconsin,
  Milwaukee, Wisconsin~53201, USA}
\author{Henri Menke}
\affiliation{Department of Physics and MacDiarmid Institute for
  Advanced Materials and Nanotechnology, University of Otago,
  P.O.~Box~56, Dunedin~9054, New Zealand}
\author{P. M. R. Brydon}
\affiliation{Department of Physics and MacDiarmid Institute for
  Advanced Materials and Nanotechnology, University of Otago,
  P.O.~Box~56, Dunedin~9054, New Zealand}
\author{Carsten Timm}
\affiliation{Institute of Theoretical Physics and W\"urzburg-Dresden
  Cluster of Excellence ct.qmat, Technische Universit\"at Dresden,
  D-01062 Dresden, Germany}
\author{Aline Ramires}
\altaffiliation[Present address: ]{Paul Scherrer Institut,
  CH-5232 Villigen, Switzerland}
\affiliation{Max Planck Institute for the Physics of Complex Systems,
  D-01187 Dresden, Germany}
\affiliation{ICTP-SAIFR, International Centre for Theoretical Physics,
  South American Institute for Fundamental Research, S\~{a}o Paulo,
  SP, 01140-070, Brazil}
\affiliation{Instituto de F\'{i}sica Te\'{o}rica, Universidade
  Estadual Paulista, S\~{a}o Paulo, SP~01140-070, Brazil}
\author{Daniel F. Agterberg}
\affiliation{Department of Physics, University of Wisconsin,
  Milwaukee, Wisconsin~53201, USA}

\date{\today}



\begin{abstract}

Strontium ruthenate (Sr$_2$RuO$_4$) has long been thought to host a spin-triplet chiral $p$-wave
superconducting state. However, the singletlike response
observed in recent spin-susceptibility measurements casts serious
doubts on this pairing state. Together with the evidence for broken
time-reversal symmetry and a jump in the shear modulus $c_{66}$
at the superconducting transition temperature, the available experiments point towards
an even-parity chiral superconductor with $k_z(k_x\pm ik_y)$-like $E_g$ symmetry,
which has consistently been dismissed based on the quasi-two-dimensional  electronic structure of
Sr$_2$RuO$_4$. Here, we show how the orbital degree of freedom can
encode the two-component nature of the $E_g$ order parameter, allowing
for a { local} or\-bi\-tal-an\-ti\-sym\-me\-tric spin-triplet state that can be stabilized by on-site
Hund's coupling. We find that this exotic $E_g$ state can be
energetically stable once a complete, realistic three-dimensional model is considered, within
which momentum-dependent spin-orbit coupling terms are key. 
This state naturally gives rise to Bogoliubov Fermi surfaces. 

\end{abstract}

\maketitle


\emph{Introduction}. Based on early Knight shift \cite{ish98},
po\-la\-rized neu\-tron scat\-ter\-ing
\cite{duf00}, muon-spin-resonance \cite{luk98}, and polar Kerr
measurements \cite{xia06}, Sr$_2$RuO$_4$ has been widely thought to
support a 
spin-triplet chiral $p$-wave superconducting state with $E_u$
symmetry
  \cite{mae94,ric95,maz99,miy99,kuw00,sat00,tak00,nom02,yan03,ann06,yos09,rag10,wan13,huo13,sca14,tsu15,mac17}.
This proposed state has had 
difficulty reconciling other experimental results \cite{mac17}, 
 including the absence of chiral edge currents
\cite{kir07}, thermal transport consistent with a nodal state
\cite{lup01,has17,kit18}, apparent Pauli-limiting effects for in-plane
fields \cite{kit14}, and the failure to observe a cusplike
behavior of the critical temperature under nematic strain \cite{hic14,ste17}.
Plausible explanations for each of these inconsistencies  have {
  nevertheless} been
presented \cite{kal09,mac17,ett18}.
Recently, however, the Knight shift has been revisited
\cite{pus19,ish19} and, contrary to earlier results, a
relatively large reduction of the Knight shift
for in-plane fields in the superconducting state  has been observed.
This  finding cannot be reconciled with the
standard spin-triplet chiral $p$-wave state~\cite{ric95}. 

Although it now seems unlikely that Sr$_2$RuO$_4$ is a 
spin-triplet chiral $p$-wave superconductor, the observation of broken time-reversal
symmetry {\cite{luk98,xia06,gri20}} and a jump in the shear modulus $c_{66}$ { \cite{lup02,gho19}}
at the critical  temperature still indicate a multicomponent
 order parameter \cite{sig91}. The only other possible
multicomponent channel within $D_{4h}$ symmetry belongs to the $E_g$  irreducible
representation (irrep) \cite{sig91}. At the Fermi surface, a chiral order
parameter in this channel resembles a spin-singlet $d$-wave state, which has horizontal line nodes. Such a state
would appear to imply that the dominant pairing instability involves
electrons in different RuO$_2$ layers, which is difficult to
understand in view of the pronounced quasi-two-dimensional nature of the normal state of Sr$_2$RuO$_4$.
Indeed, no microscopic calculation for Sr$_2$RuO$_4$  has found a leading weak-coupling $E_g$ instability~\cite{zut05,sch06,roi19}. 

In this { Rapid Communication}, we show that \emph{local} interactions can lead to a
weak-coupling instability in the $E_g$ channel, once we consider a
complete three-dimensional (3D) model for the normal
state. Physically, this $E_g$ state is a { local (i.e., $s$-wave)}
or\-bi\-tal-an\-ti\-sym\-me\-tric spin-triplet  (OAST) state stabilized
by on-site Hund's coupling. When the
renormalized low-energy Hund's coupling $J$ becomes larger than the
interorbital Hubbard interaction $U'$, this  channel develops
an attractive interaction \cite{pue12,spa01,han04,vaf17,che19,lind19}.  This pairing instability has been found in dynamical mean-field theory, which predicts it appears in the strong-coupling limit even when the unrenormalized high-energy $J$ is less than $U'$ \cite{hos15}, and also  in the presence of strong charge fluctuations \cite{gin18}. Pairing due to this type of interaction was considered for Sr$_2$RuO$_4$ in Ref.\ \cite{pue12} , where an $A_{1g}$
pairing state was found to be stable. Motivated by the relevance
of $J$ for the normal state of Sr$_2$RuO$_4$ \cite{tam19},
we revisit the local-pairing scenario.
{ We note that, remarkably, a similar OAST pairing state is believed to be responsible for nematic superconductivity in Cu$_x$Bi$_2$Se$_3$ \cite{fu10,sas11,yam12}.}
In the following, we show that an $E_g$ state can be stabilized over the $A_{1g}$ state  of Ref.\ \cite{pue12} by including  momentum-dependent spin-orbit coupling (SOC) corresponding to interlayer spin-dependent hopping with  a hopping integral on the order of $10\,\mathrm{meV}$. This small value leaves the quasi-two-dimensional nature of the band structure intact.  Moreover, we use the concept of superconducting fitness \cite{ram16,ram18} to understand the importance of this term in stabilizing the $E_g$ state. Finally, we show that this chiral multiorbital $E_g$ state will display Bogoliubov Fermi surfaces \cite{agt17,bry18}, instead of line nodes.


\emph{Normal-state Hamiltonian}. An accurate description of the
normal-state Hamiltonian is crucial for understanding
superconductivity in the weak-coupling limit. Our starting point is a tight-binding
parametrization of the normal-state Hamiltonian
that includes all terms
allowed by symmetry \cite{ram19}. To determine the magnitude of each term, we carry out a  fit to the density-functional theory (DFT) results of Veenstra \emph{et al.}\
\cite{vee14}. Details on the numerical procedures are
provided in the Supplemental Material (SM) \cite{SM}\nocite{SM:Gradhand2013, SM:Huang2018, SM:dlib}. However,
angle-resolved photoemission spectroscopy (ARPES) measurements
\cite{tam19,zhang16} suggest that some DFT parameters differ appreciably
from the measured values, in particular the SOC  strengths \cite{vee14}.
We therefore allow the  SOC parameters to
vary in order to understand how  they affect the  
leading superconducting instability, under the constraint that the
Fermi surfaces do not differ significantly from the DFT predictions
and are hence still qualitatively in accordance
with the ARPES results.  

The relevant low-energy degrees of freedom (DOF) are the electrons in the
$t_{2g}$-orbital manifold  $d_{yz}$, $d_{xz}$, and $d_{xy}$ of Ru. Using the spinor operator
$\Phi^\dagger_{\bk} = (c_{\bk,yz\up}^\dagger,\allowbreak c_{\bk,yz\dn}^\dagger,\allowbreak
c_{\bk,xz\up}^\dagger,\allowbreak c_{\bk,xz\dn}^\dagger,\allowbreak
c_{\bk,xy\up}^\dagger,\allowbreak c_{\bk,xy\dn}^\dagger)$, where
$c^\dagger_{\bk,\gamma\sigma}$ creates an electron with momentum $\bk$
and spin $\sigma$ in orbital $\gamma$, we 
construct the most general three-orbital single-particle Hamiltonian  as
$H_0=\sum_{\bk}\Phi^\dagger_\bk \hat{H}_0(\bk) \Phi_\bk$ with
\begin{equation}\label{Eq:H0}
\hat{H}_0(\bk) = \sum_{a=0}^8\sum_{b=0}^3 h_{ab}(\bk)\, \lambda_a \otimes \sigma_b,
\end{equation}
where  the $\lambda_a$ are Gell-Mann matrices encoding the orbital DOF and the
$\sigma_b$ are Pauli matrices encoding the  spin
($\lambda_0$ and $\sigma_0$ are unit matrices), and $h_{ab}(\bk)$
are even functions of  momentum. Time-reversal and
inversion symmetries allow only for 15 $h_{ab}(\bk)$
functions to be finite. The explicit form of the $h_{ab}(\bk)$ functions and the Gell-Mann matrices are  given in the SM~\cite{SM}.


\emph{Interactions and superconductivity}. We consider on-site interactions of the Hubbard-Kanamori type~\cite{dag01},
\begin{align}
H_\text{int} &= \frac{U}{2} \sum_{i,\gamma,\sigma\neq\sigma^\prime} n_{i\gamma\sigma} n_{i\gamma\sigma^\prime}
  + \frac{U^\prime}{2} \sum_{ i,\gamma\neq\gamma^\prime,\sigma,\sigma^\prime}
    n_{i\gamma\sigma}n_{i\gamma^\prime\sigma^\prime} \nonumber \\
&\quad{}+ \frac{J}{2} \sum_{ i,\gamma\neq\gamma^\prime,\sigma,\sigma^\prime} c_{i\gamma\sigma}^\dag
  c_{i\gamma^\prime \sigma^\prime}^\dag c_{i\gamma\sigma^\prime} c_{i\gamma^\prime\sigma} \nonumber \\
&\quad{}+  \frac{J^{\prime}}{2} \sum_{ i,\gamma\neq\gamma^\prime,\sigma\neq\sigma^\prime} c_{i\gamma\sigma}^\dag
  c_{i\gamma\sigma^\prime}^\dag c_{i\gamma^\prime\sigma^\prime} c_{i\gamma^\prime\sigma},
\label{Eq:Int}
\end{align}
where $c_{i\gamma\sigma}^\dagger$ ($c_{i\gamma\sigma})$ creates
(annihilates) an electron at site $i$ in orbital $\gamma$ with spin
$\sigma$, and $n_{i\gamma\sigma} = c_{i\gamma\sigma}^\dagger
c_{i\gamma\sigma}$. The first two terms describe repulsion
($U,U^\prime>0$) between electrons in the same and in different orbitals, respectively. The third and fourth terms represent the Hund's
exchange interaction and pair-hopping interactions respectively. We take $J=J^{\prime}$ \cite{dag01}, where $J>0$ is expected for Sr$_2$RuO$_4$. 
In the context of Sr$_2$RuO$_4$, $H_\text{int}$
is usually taken as the starting point for the calculation of the spin- and
charge-fluctuation propagators which enter into the effective
 interaction \cite{rag10,sca14}. 
Here, we take a different approach \cite{pue12,vaf17}  by directly
decoupling the interaction in the Cooper channel, which, for $U'-J<0$,
yields an attractive interaction for on-site pairing  in an OAST state. This scenario has previously been
applied to a  two-dimensional  model of
Sr$_2$RuO$_4$, predicting an  OAST $A_{1g}$
state \cite{pue12}. Although a strong-coupling instability towards
an OAST $E_g$ state in the absence of SOC has
been predicted in Ref.\ \cite{gin18}, the superconductivity in
Sr$_2$RuO$_4$ is likely  in the weak-coupling regime
\cite{mac17}. It is therefore important to understand if an
OAST $E_g$ state can be the leading
instability in this limit.

\begin{table}
  \centering
  \begin{ruledtabular}
    \begin{tabular}{@{}ccccc@{}}
      Irrep                     & $[a,b]$               & Orbital       & Spin    & Interaction  $g$ \\
      \colrule
      \multirow{4}{*}{$A_{1g}$} & $[0,0]$               & symmetric     & singlet & $U+2J$           \\
                                & $[8,0]$               & symmetric     & singlet & $U-J$            \\
                                & $[4,3]$               & antisymmetric & triplet & $U'-J$           \\
                                & $[5,2]-[6,1]$         & antisymmetric & triplet & $U'-J$           \\
      \colrule
      $A_{2g}$                  & $[5,1]+[6,2]$         & antisymmetric & triplet & $U'-J$           \\
      \colrule
      \multirow{2}{*}{$B_{1g}$} & $[7,0]$               & symmetric     & singlet & $U-J$            \\
                                & $[5,2]+[6,1]$         & antisymmetric & triplet & $U'-J$           \\
      \colrule
      \multirow{2}{*}{$B_{2g}$} & $[1,0]$               & symmetric     & singlet & $U'+J$           \\
                                & $[5,1]-[6,2]$         & antisymmetric & triplet & $U'-J$           \\
      \colrule
      \multirow{3}{*}{$E_g$}    & { $\{[2,0],[3,0]\}$}  & symmetric     & singlet & $U'+J$           \\
                                & { $\{[4,1],[4,2]\}$}  & antisymmetric & triplet & $U'-J$           \\
                                & { $\{[6,3],-[5,3]\}$} & antisymmetric & triplet & $U'-J$           \\
    \end{tabular}
  \end{ruledtabular}
\caption{All  even-parity local gap functions classified by  irreps of
  the point group $D_{4h}$. Here, $[a,b]$  corresponds to the
  parametrization of the gap matrix as $\lambda_a \otimes \sigma_b\,
  (i\sigma_2)$. The  other columns give the orbital and spin
  character, as well as the interaction $g$ for each superconducting
  state derived from the Hubbard-Kanamori interaction  $H_\text{int}$
  in Eq.\ (\ref{Eq:Int}). Note that the  two components of the $E_g$
  order parameters can stem from the orbital DOF, as for {
    $\{[2,0],[3,0]\}$} and { $\{[6,3],-[5,3]\}$}, or from the spin
  DOF, as for { $\{[4,1],[4,2]\}$}.}
\label{Tab:Gap}
\end{table}

In the spirit of Ref.\ \cite{vaf17}, we treat  $H_\text{int}$ as a renormalized low-energy effective 
interaction. We tabulate the allowed local gap functions, their symmetries, and the  interactions in the respective  pairing channels in Table \ref{Tab:Gap}.
Here, we adopt the common assumption of
on-site rotational symmetry, which stipulates
$U=U^\prime+2J$ \cite{dag01}. This choice
implies that all the  OAST channels have
the same attractive pairing interaction,  which highlights the role of the normal-state Hamiltonian in selecting the most stable state.
 However, since the Ru sites have $D_{4h}$ symmetry and not the assumed  full rotational symmetry, the interaction strengths for the
different pairing channels are generally different. Our results
should therefore be interpreted as providing a guide to which superconducting states this
form of attractive  interaction can give rise to.   

We write a free-energy expansion up to second order in the superconducting order parameter given by the gap matrices $\hat{\Delta}_i=\Delta_i\, \lambda_{a_i} \otimes \sigma_{b_i} (i\sigma_2)$,
\begin{equation}\label{eq:gap}
\mathcal{F} = \frac{1}{2}\sum_{i}
              \frac{1}{g_i}\, \textrm{Tr}\, [\hat{\Delta}_i^\dag \hat{\Delta}_i]
              - \frac{k_B T}{2} \sum_{\bk,\omega,i,j}
                \textrm{Tr}\, \big[\hat{\Delta}_i \hat{\underline{G}} \hat{\Delta}^\dag_j \hat{G}\big],
\end{equation}
where $i$ and $j$ sum over all channels of a chosen irrep, $g_i$ are the corresponding interaction  strengths from Table \ref{Tab:Gap}, $\omega_m = (2m+1)\pi k_B T$
are the fermionic Matsubara frequencies, and
$\hat{G}=(i\omega_m - \hat{H}_0)^{-1}$ and $\hat{\underline{G}}=(i\omega_m + \hat{H}^T_0)^{-1}$ are the normal-state Green's functions. Nontrivial solutions of
the coupled linearized gap equations  obtained from $\partial \mathcal{F}/\partial
\Delta_i^\ast = 0$ give the critical temperature $T_c$ and the
 linear combination of  the $\hat{\Delta}_i$ corresponding to the leading instability.
 We include all channels in a chosen irrep, not just the attractive ones (see Table \ref{Tab:Gap}).
In evaluating the  last term in Eq.\ (\ref{eq:gap}), we keep only
intraband contributions; although the inclusion of interband terms
will shift  $T_c$, this effect is negligible in the weak-coupling regime, as discussed in detail in the SM~\cite{SM}.


\emph{Results}. Weak-coupling OAST pairing states for an attractive Hund's interaction require nonvanishing SOC \cite{vaf17,ram18,che19}. As described in the SM \cite{SM}, SOC
appears in five  terms in the Hamiltonian $\hat{H}_0(\bk)$ in Eq.\ (\ref{Eq:H0}), representing a large parameter space to
explore. We shall focus on the effects of the following  terms:
the $z$ component of the atomic SOC, $h_{43} = \eta_z$;
the in-plane atomic SOC, $h_{52}-h_{61}=\eta_\perp$;
and the momentum-dependent SOC associated with the interlayer hopping amplitude
$t^{\text{SOC}}_{56z}$ between the $d_{xy}$ and the $d_{xz}$ and $d_{yz}$ orbitals,
$\{h_{53},h_{63}\} = 8\, t^{\text{SOC}}_{56z}\, \sin(k_zc/2)\allowbreak \{ \cos(k_xa/2)\allowbreak\sin(k_ya/2),\allowbreak {-}\sin(k_xa/2)\allowbreak\cos(k_ya/2) \}$.
Here, we will ignore the anisotropy of the atomic SOC and set $\eta_z = \eta_\perp = \eta$.
We have carried out a cursory exploration of the larger SOC parameter space and find that varying
the other parameters within reasonable ranges such that the Fermi
surfaces do not significantly deviate from the DFT predictions has
little effect on the leading instability.

\begin{figure}
\begin{center}
	\includegraphics[width=\columnwidth]{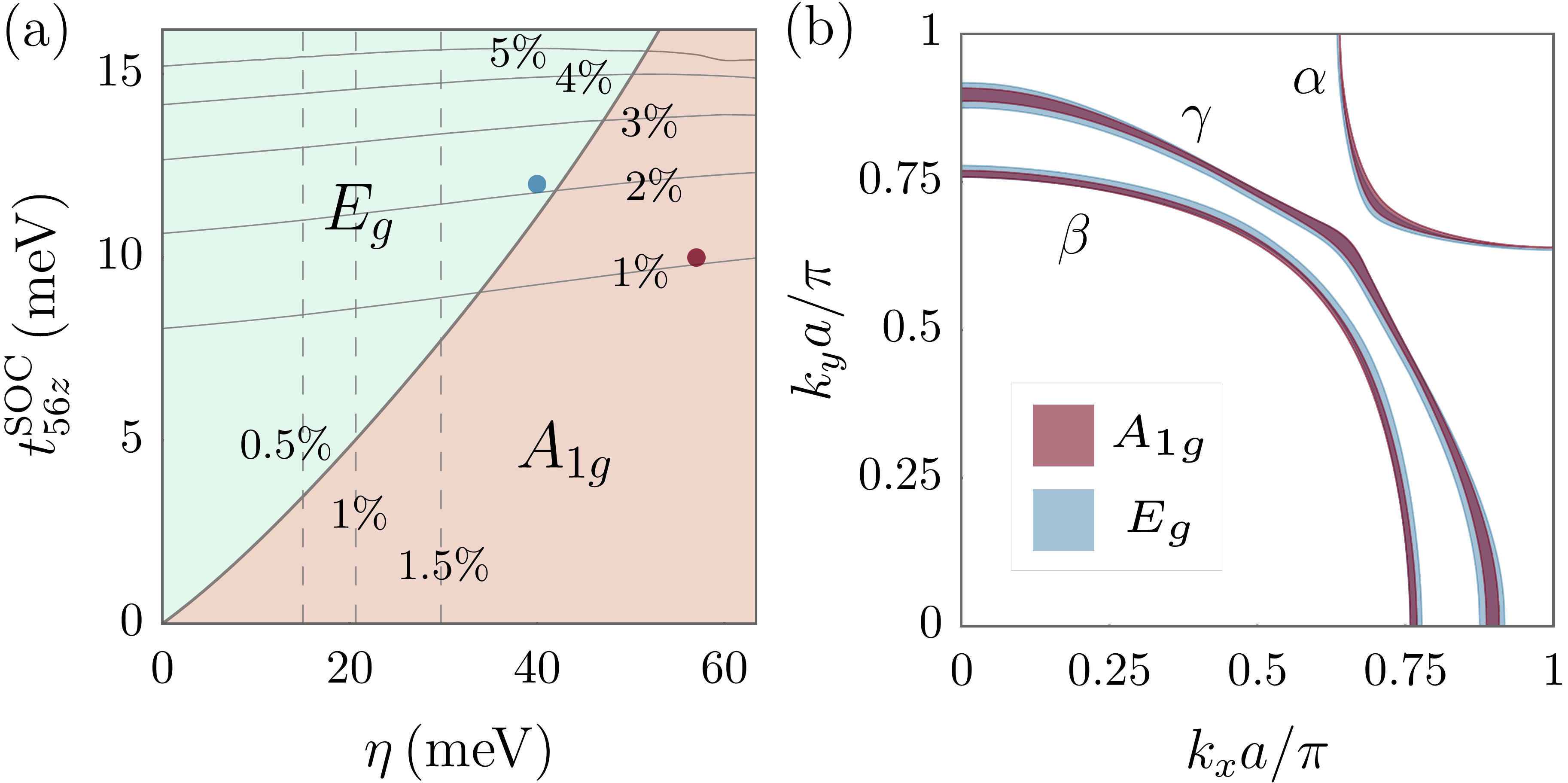}
\end{center}
	\caption{
(a)~Phase diagram  showing the stability of $A_{1g}$ and $E_g$ pairing states as a function of the SOC parameters  $\eta$ and $t_{56z}^\text{SOC}$. The vertical dashed lines indicate the minimum distance between two Fermi surfaces. Percentages are defined as fractions of $2\pi/a$.
For small $\eta$,  the separation between the $\beta$ and $\gamma$ bands becomes too small, in view of the ARPES data \cite{tam19}.
The  thin solid lines indicate the maximum  variation of the Fermi surface along the $k_z$ direction. For large $t^\text{SOC}_{56z}$, the Fermi surfaces become too  dispersive. The blue and magenta dots denote the parameter choices for $E_g$ and $A_{1g}$ stable solutions used in (b).
(b)~Fermi-surface shapes, projected onto the $k_xk_y$ plane, for representative points in the $A_{1g}$ (red) and $E_g$ (blue) regions in (a). For $A_{1g}$,
 $\eta = 57\, \mathrm{meV}$ and $t^\text{SOC}_{56z} = 10\, \mathrm{meV}$, while for $E_g$,
$\eta = 40\, \mathrm{meV}$ and $t^\text{SOC}_{56z} = 12\, \mathrm{meV}$.
}\label{Fig:PD}
\end{figure}

Figure \ref{Fig:PD}(a) shows the phase diagram as a function of the atomic SOC $\eta$ and the momentum-dependent SOC, parametrized by $t^{\text{SOC}}_{56z}$. We find leading instabilities in the $A_{1g}$ and $E_g$ channels.  $A_{2g}$ and $B_{2g}$ states are not competitive anywhere in the phase diagram.  A $B_{1g}$ state is sometimes found as a subleading instability.
The $E_g$ solution
is dominated by the { $\{[6,3],-[5,3]\}$} channel and is stabilized for $t^{\text{SOC}}_{56z}\gtrsim \eta/4$. Under the
constraint of realistic Fermi surfaces, the $E_g$ state can be
stabilized for $t^{\text{SOC}}_{56z}$ as small as about $5\,\mathrm{meV}$,
although this requires a rather small value  of the on-site SOC. It is 
remarkable that such a small energy scale determines the relative
stability of qualitatively different pairing states. As shown in
Fig.~\ref{Fig:PD}(b), the Fermi
surfaces for parameters stabilizing $A_{1g}$ or $E_g$
states are indeed very similar. The SOC strength remains controversial~\cite{vee14,tam19,zhang16}, but here we have shown its importance for the determination of the most stable superconducting state.
Our results are a proof of principle that an $E_g$
superconducting state can be realized in Sr$_2$RuO$_4$,
even for purely local interactions, once one properly takes
into account a complete and plausible 3D model for the normal state.

\begin{figure}[t]
  \centering
  \includegraphics[width=\columnwidth]{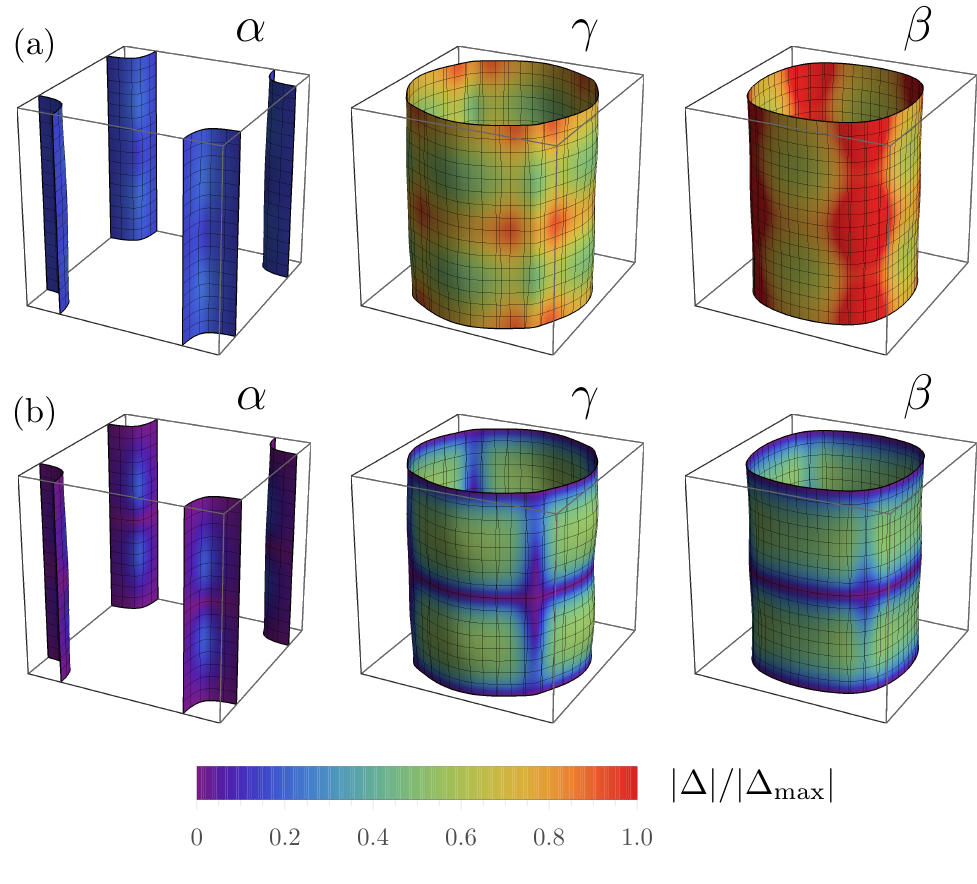}
  \caption{Projected gaps at the Fermi surfaces for a representative
    (a) $A_{1g}$ and (b) { chiral} $E_g$ state in the first Brillouin zone. Parameters are the same as in Fig. 1 (b). 
    The color code is
    normalized to the maximum value of the $A_{1g}$ gap.}
  \label{Fig:FS}
\end{figure}

Figure~\ref{Fig:FS} displays the projected gaps at the Fermi surfaces for representative $A_{1g}$ and $E_g$ states. Note that in both cases the gap magnitude  on the $\alpha$ sheet is very small, whereas the gaps  on the $\beta$ and $\gamma$ sheets are comparable. This  shows that we cannot simply identify the $\gamma$ band \cite{agt97} or the pair of almost one-dimensional $\alpha$ and $\beta$ bands \cite{rag10} as the dominant  ones for superconductivity \cite{sca14}.

It is possible to  understand why these SOC terms
stabilize the respective ground states based on the notion of
superconducting fitness \cite{ram16,ram18}. In particular, it has been
shown for two-band superconductors that if the quantity
 { $\hat F_A(\bk)=\tilde{H}_0(\bk)\hat{\Delta}(\bk)+\hat{\Delta}(\bk)\tilde{H}_0^\ast(-\bk)$}
is zero there is no intraband pairing and hence no weak-coupling instability [here,
$\tilde{H}_0(\bk)$ corresponds to $\hat{H}_0(\bk)$ with  $h_{00}(\bk)$ set to zero].
Hence, adding terms to the normal-state Hamiltonian such that
$\hat F_A(\bk)$ becomes nonzero for a particular gap function turns
on a weak-coupling instability in this channel. The fitness analysis can be extended to our three-orbital model or, alternatively, we can construct an effective two-orbital model valid sufficiently far from the Brillouin-zone diagonals. Applying
the fitness argument to the effective two-band model, we find that the on-site
SOC $\eta$ turns on both the $A_{1g}$ and $B_{1g}$ pairing channels,
whereas the parameter $t^{\text{SOC}}_{56z}$ turns on the $E_g$
 $\{[6,3],-[5,3]\}$ channel, consistent with what we find
numerically.  Details of the fitness analysis are given in the SM~\cite{SM}.

In view of the Knight-shift experiments \cite{pus19,ish19},
it is important to comment on the
spin susceptibility associated with the dominant $E_g$
$\{[6,3],-[5,3]\}$ channel. Since it is a spin-triplet
state with in-plane spin polarization of the Copper pairs, similar to the familiar chiral $p$-wave spin-triplet pairing with a ${\bm d}$-vector along the $k_z$ direction, it might naively be
expected to show a temperature-independent spin susceptibility
for in-plane fields. This is not the case, however, since the even parity of $E_g$ implies that the intraband pairing potential is a pseudo-spin singlet when
expressed in the band basis and the low-energy response
to a magnetic field is identical to a true spin singlet. This has been examined numerically for similar pairing states \cite{yu18,lind19}, where it was found that only a small fraction of the normal-state spin susceptibility persists at zero temperature in the superconducting state.

\begin{figure}
\begin{center}
\includegraphics[width=\columnwidth]{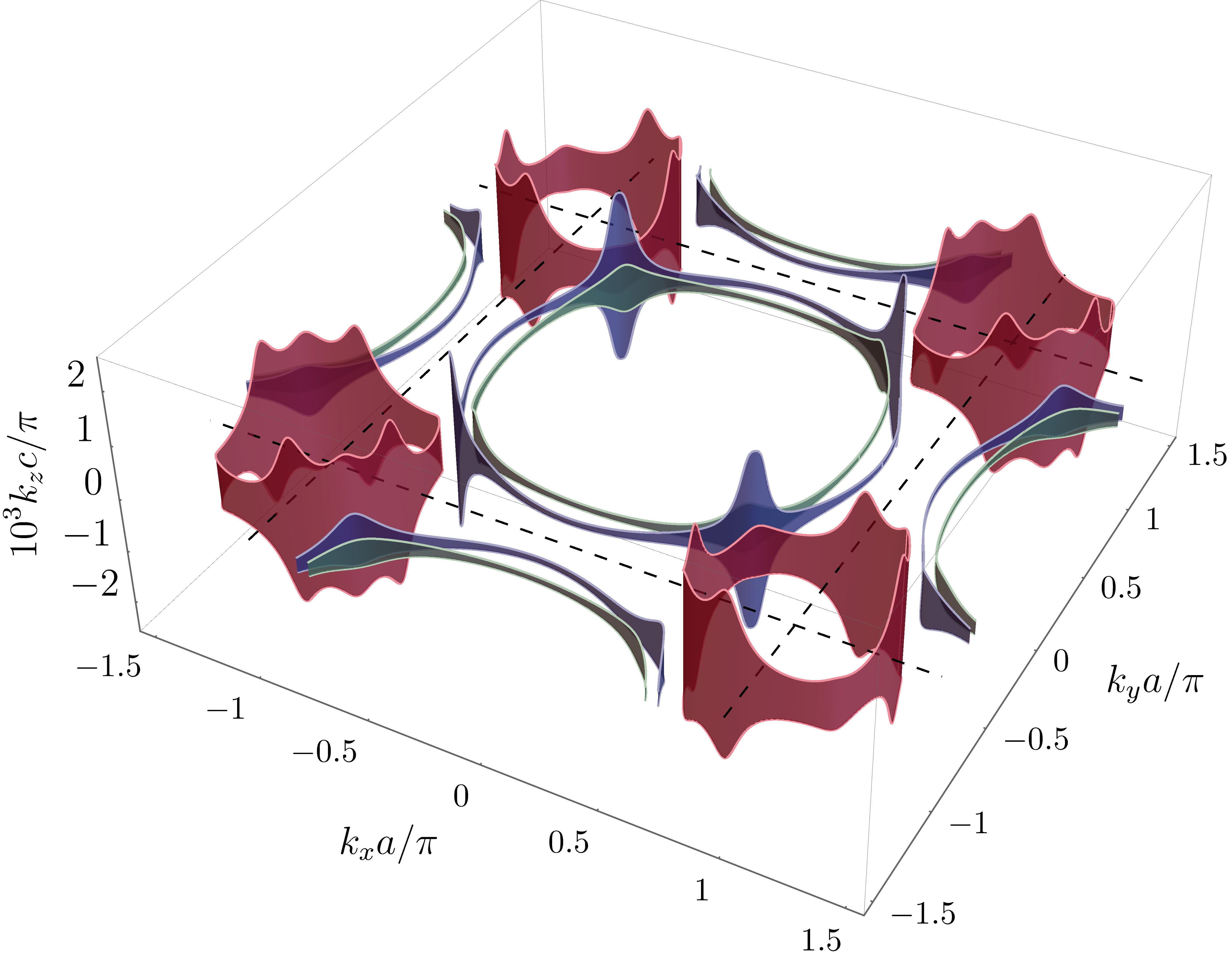}
\caption{ BFSs for the chiral $E_g$ state. The Fermi surfaces in red, green, and blue correspond to inflated nodes stemming from the $\alpha$, $\beta$, and $\gamma$  band, respectively.
}
\label{Fig:BFS}
\end{center}
\end{figure}


\emph{Bogoliubov Fermi surfaces}. An $E_g$ state is expected
to have horizontal line nodes  at $k_z=0$ and  $2\pi/c$
\cite{sig91,mac17}, and it { will} have vertical line
  nodes in a time-reversal invariant nematic state
  \cite{sig91,lup02}. { Although recent tunneling measurements have
  called into question time-reversal symmetry breaking in
  Sr$_2$RuO$_4$~\cite{kas19}, here we follow the indications of polar Kerr
and $\mu$SR experiments~\cite{luk98,xia06,gri20}, and} explicitly consider a chiral $E_g$ state
which  has no vertical line nodes. It has recently been shown that
for an even-parity superconductor that spontaneously breaks time-reversal
symmetry, the excitation spectrum is either fully gapped or contains
 Bogoliubov Fermi surfaces  (BFSs) \cite{agt17,bry18}. Indeed,
 the chiral $E_g$ state considered here has BFSs, which are shown in
Fig.\ \ref{Fig:BFS}. These  BFSs are very thin in the
direction perpendicular to the normal-state Fermi surface, giving them a
ribbon-like appearance that extends along the $k_z$ axis 
by  about  0.2\% of the
Brillouin zone. This value is proportional to the gap  amplitude, here set to $0.15\,\mathrm{meV}$. While the total residual density
of states from the  BFSs is not large and may be difficult to observe \cite{lapp19},
such a nodal structure implies that some experimental results require
reinterpretation. In particular, given that the  BFSs extend
along the $k_z$-axis, the  argument that thermal
conductivity measurements rule out the  $E_g$ state because it has horizontal line nodes \cite{has17}
no longer applies. The presence of BFSs may also require a
reinterpretation of quasi-particle-interference experiments
\cite{shar19}. We leave a detailed study of experimental consequences
of the $E_g$ OAST state for future work. 


\emph{Conclusions}. We have argued that an $E_g$ order parameter can
be a realistic weak-coupling ground state for Sr$_2$RuO$_4$, once we
consider a complete 3D model for the normal state and interactions of
the Hubbard-Kanamori type. Key to our construction are the usually
neglected  momentum-dependent SOC terms in the normal state. These
terms can completely change the nature of the superconducting ground
state, despite being so small that  they do not  significantly change
the Fermi surfaces. Our theory reconciles
the recent observation of a singlet-like spin susceptibility with
measurements indicating a two-component order parameter and broken
time-reversal symmetry. 


\emph{Acknowledgments}. The authors thank  A. V. Chu\-bu\-kov and B. Ramshaw for
useful discussions. H.M. and P.M.R.B. were supported by the Marsden Fund
Council from Government funding, managed by Royal Society Te
Ap\={a}rangi. A.R. acknowledges the support of Funda\c{c}\~{a}o de Amparo \`{a} Pesquisa do Estado de S\~{a}o Paulo (FAPESP) Project No.~2018/18287-8, and Funda\c{c}\~{a}o  para o Desenvolvimento da UNESP (FUNDUNESP) Process No.~2338/2014-CCP. C.T. acknowledges financial support by the Deutsche 
Forschungsgemeinschaft through the Collaborative Research Center SFB 
1143, Project A04, the Research Training Group GRK 1621, and the Cluster 
of Excellence on Complexity and Topology in Quantum Matter ct.qmat (EXC 
2147).

H.G.S. and H.M. contributed equally to this work.

\newcommand\enquote{}
\bibliography{references}

\onecolumngrid

\newpage

\renewcommand{\theequation}{S\arabic{equation}}
\renewcommand{\thefigure}{S\arabic{figure}}
\renewcommand{\thetable}{S\arabic{table}}
\setcounter{page}{1}
\setcounter{equation}{0}
\setcounter{figure}{0}
\setcounter{table}{0}

\begin{center}
\textbf{{\large Supplemental Material for\\[0.5ex]
Stabilizing Even-Parity Chiral Superconductivity in Sr$_2$RuO$_4$}}\\[1.5ex]
Han Gyeol Suh, Henri Menke, P. M. R. Brydon, Carsten Timm, Aline Ramires, and Daniel F. Agterberg
\end{center}

\section{Microscopic model}

In this section, we construct a 3D tight-binding model for
Sr$_2$RuO$_4$. We take into account the full 3D Fermi surfaces (FSs)
of Sr$_2$RuO$_4$, based on the DFT band structure obtained by Veenstra
\textit{et al.}~\citeS{S:Veenstra2014}, who showed that despite the 2D
shape of the FSs, the orbital and spin polarization vary along $k_z$.
To account for the presence of orbital mixing on the different FS
sheets, we include the $t_{2g}$ manifold of the Ru $d_{yz}$, $d_{xz}$,
and $d_{xy}$ orbitals (we will assume this order throughout).

We parametrize the orbital space by the the Gell-Mann matrices, which
are the generators of $\mathrm{SU}(3)$. We use the convention
\begin{equation}
  \renewcommand\arraystretch{1}%
  \begin{aligned}
    \lambda_0 &= \begin{pmatrix} 1 & 0 & 0 \\ 0 & 1 & 0 \\ 0 & 0 & 1 \end{pmatrix}, &
    \lambda_1 &= \begin{pmatrix} 0 & 1 & 0 \\ 1 & 0 & 0 \\ 0 & 0 & 0 \end{pmatrix}, &
    \lambda_2 &= \begin{pmatrix} 0 & 0 & 1 \\ 0 & 0 & 0 \\ 1 & 0 & 0 \end{pmatrix}, &
    \lambda_3 &= \begin{pmatrix} 0 & 0 & 0 \\ 0 & 0 & 1 \\ 0 & 1 & 0 \end{pmatrix}, &
    \lambda_4 &= \begin{pmatrix} 0 & -i & 0 \\ i & 0 & 0 \\ 0 & 0 & 0 \end{pmatrix}, \\
    &&
    \lambda_5 &= \begin{pmatrix} 0 & 0 & -i \\ 0 & 0 & 0 \\ i & 0 & 0 \end{pmatrix}, &
    \lambda_6 &= \begin{pmatrix} 0 & 0 & 0 \\ 0 & 0 & -i \\ 0 & i & 0 \end{pmatrix}, &
    \lambda_7 &= \begin{pmatrix} 1 & 0 & 0 \\ 0 & -1 & 0 \\ 0 & 0 & 0 \end{pmatrix}, &
    \lambda_8 &= \frac{1}{\sqrt{3}} \begin{pmatrix} 1 & 0 & 0 \\ 0 & 1 & 0 \\ 0 & 0 & -2 \end{pmatrix}.
  \end{aligned}
\end{equation}
We write the normal-state Hamiltonian in terms of the spinor
$\Phi_{\bm{k}} = (c_{\bm{k},2,\up}, c_{\bm{k},2,\dn},
c_{\bm{k},1,\up}, c_{\bm{k},1,\dn}, c_{\bm{k},3,\up},
c_{\bm{k},3,\dn})^T$, where we introduce the numbering of the orbitals
$1 = d_{xz}$, $2 = d_{yz}$, $3 = d_{xy}$. In terms of the Gell-Mann
and Pauli matrices, we write the Hamiltonian
$H_0 = \sum_{\bm{k}} \Phi_{\bm{k}}^\dagger \hat{H}_0(\bm{k})
\Phi_{\bm{k}}$ where
\begin{equation}
  \label{S:Eq:H0}
  \hat{H}_0(\bm{k}) = \sum_{a,b} h_{ab}(\bm{k})\, \lambda_a \otimes \sigma_b .
\end{equation}
In the presence of inversion and time-reversal symmetries, only a
subset of fifteen $h_{ab}(\bm{k})$ terms are
allowed. Table~\ref{Tab:H0} lists the symmetry-allowed terms, the
associated irrep for the matrices $\lambda_a \otimes \sigma_b$, the
physical process to which these correspond, and their momentum
dependence.

\begin{table}[ht]
\begin{center}
  \renewcommand\arraystretch{1.25}
    \begin{tabular}{| c | c | c | c | c |}
      \hline
      Irrep & $(a,b)$ & Type & Explicit form of $h_{ab}(\bk)$ \\
      \hline \hline
      \multirow{4}{*}{$A_{1g}$} & $(0,0)$ & intra-orbital hopping & $\frac{1}{3}\, [\xi_{11}(\bm{k}) + \xi_{22}(\bm{k}) + \xi_{33}(\bm{k})]$ \\
      \cline{2-4}
      & $(8,0)$ & intra-orbital hopping & $\frac{1}{2\sqrt{3}}\, [\xi_{11}(\bm{k}) + \xi_{22}(\bm{k}) - 2 \xi_{33}(\bm{k})]$ \\
      \cline{2-4}
      & $(4,3)$ & atomic SOC & $-\eta_z$ \\
      \cline{2-4}
      & $(5,2)-(6,1)$ & atomic-SOC & $\eta_\perp$ \\
      \hline
      $A_{2g}$ & $(5,1)+(6,2)$ & $\bk$-SOC & neglected \\
      \hline
      \multirow{2}{*}{$B_{1g}$} & $(7,0)$ & intra-orbital hopping & $\frac{1}{2}\, [\xi_{22}(\bm{k}) - \xi_{11}(\bm{k})]$ \\
      \cline{2-4}
      & $(5,2)+(6,1)$ & $\bk$-SOC & $2 t^{\mathrm{SOC}}_{5261} (\cos k_xa - \cos k_ya)$ \\
      \hline
      \multirow{2}{*}{$B_{2g}$} & $(1,0)$ & inter-orbital hopping & $\lambda(\bm{k})$ \\
      \cline{2-4}
      & $(5,1)-(6,2)$ & $\bk$-SOC & $4 t^{\mathrm{SOC}}_{5162} \sin k_xa \sin k_ya$ \\
      \hline
      \multirow{3}{*}{$E_g$} & $\{(2,0),(3,0)\}$ & inter-orbital hopping & $8 t^{13}_z \sin(k_zc/2) \{ \sin(k_xa/2) \cos(k_ya/2),\cos(k_xa/2) \sin(k_ya/2) \}$ \\
      \cline{2-4}
      & $\{(4,1),(4,2)\}$ & $\bk$-SOC & $8 t^{\mathrm{SOC}}_{12z} \sin(k_zc/2) \{ \sin(k_xa/2) \cos(k_ya/2),\cos(k_xa/2) \sin(k_ya/2)\}$ \\
      \cline{2-4} 
      & $\{(6,3),-(5,3)\}$ & $\bk$-SOC & $-8 t^{\mathrm{SOC}}_{56z} \sin(k_zc/2) \{\sin(k_xa/2) \cos(k_ya/2),\cos(k_xa/2) \sin(k_ya/2) \}$ \\
      \hline
    \end{tabular}
\end{center}
\caption{List of the fifteen symmetry-allowed terms in the
  normal-state Hamiltonian $\hat{H}_0(\bk)$ in
  Eq.~(\ref{S:Eq:H0}). For each $(a,b)$, the basis function
  $h_{ab}(\bk)$ must belong to the same irrep of $D_{4h}$ as the
  matrix $\lambda_a \otimes \sigma_b$. The table gives the irrep, the
  associated physical process (``Type''), where ``$\bk$-SOC'' means
  momentum-dependent (nonlocal) SOC, and the momentum dependence of
  $h_{ab}(\bk)$. For the two-dimensional irrep $E_g$, the entries are
  organized such that the first transforms as $xz$ and the second as
  $yz$.}
\label{Tab:H0}
\end{table}

\goodbreak Note that Table~\ref{Tab:H0} has entries which are in
accordance with previous literature~\citeS{S:Gradhand2013,
  S:Scaffidi2014, S:Huang2018} but there are also new terms associated
with hopping along the $z$-direction or momentum-dependent SOC, which
are usually neglected. Here we take $\eta_z = \eta_\perp = \eta$ as the parameter for the on-site atomic SOC. The intra-orbital hoppings $\xi_{11,22,33}(\bk)$ are
included up to next-next-nearest neighbors in plane and next-nearest
neighbors out of plane. The inter-orbital hopping $\lambda(\bk)$
between the $d_{xz}$ and the $d_{yz}$ orbitals is kept up to
next-nearest neighbors in plane and nearest neighbors out of
plane. For the inter-orbital hopping $\{(3,0), -(2,0)\}$ between the
$d_{xz}$ and $d_{xy}$ ($d_{yz}$ and $d_{xy}$) orbitals, we only keep
the nearest-neighbor component out of plane. The explicit form of the
functions not given explicitly in Table~\ref{Tab:H0} is
\begin{align}
   \xi_{11,22}(\bm k) ={}& 2 t^{11}_{x,y} \cos k_xa + 2 t^{11}_{y,x} \cos k_ya \nonumber \\
      &{} + 8 t^{11}_z \cos(k_xa/2) \cos(k_ya/2) \cos(k_zc/2) \nonumber \\
      &{} + 4 t^{11}_{xy} \cos k_xa \cos k_ya + 2 t^{11}_{xx,yy} \cos 2 k_xa + 2 t^{11}_{yy,xx} \cos 2 k_ya
        \nonumber \\
      &{} + 4 t^{11}_{xxy,xyy} \cos 2 k_xa \cos k_ya + 4 t^{11}_{xyy,xxy} \cos 2 k_ya \cos k_xa \nonumber \\
      &{} + 2 t^{11}_{zz} (\cos k_zc - 1) - \mu, \\
  \xi_{33}(\bm k) ={}& 2 t^{33}_{x} (\cos k_xa + \cos k_ya) \nonumber \\
      &{} + 8 t^{33}_z \cos(k_xa/2) \cos(k_ya/2) \cos(k_zc/2) \nonumber \\
      &{} + 4 t^{33}_{xy} \cos k_xa \cos k_ya + 2 t^{33}_{xx} (\cos 2 k_xa + \cos 2 k_ya) \nonumber \\
      &{} + 4 t^{33}_{xxy} (\cos 2 k_xa \cos k_ya + \cos 2 k_ya \cos k_xa) \nonumber \\
      &{} + 2 t^{33}_{zz} (\cos k_zc - 1) - \mu_1, \\
  \lambda(\bm k) ={}& 4 t^{12}_z \sin(k_xa/2) \sin(k_ya/2) \cos(k_zc/2) \nonumber \\
      &{} - 4 t^{12}_{xy} \sin k_xa \sin k_ya \nonumber \\
      &{} - 4 t^{12}_{xxy} (\sin 2 k_xa \sin k_ya + \sin 2 k_ya \sin k_xa) .
\end{align}
We now focus on terms corresponding to $\bk$-dependent SOC, usually
not taken into account in the standard parametrization of the
normal-state Hamiltonian. The first matrix in the list,
$\lambda_5\otimes\sigma_1 + \lambda_6\otimes\sigma_2$, which is of
$A_{2g}$ symmetry, will be ignored because the lowest-order polynomial
basis function of this irrep is of order 4 ($g$-wave), which only
appears at next-next-next-nearest-neighbor hopping and is therefore
assumed to be negligible. We also take the other $\bk$-dependent SOC
terms at the lowest order at which they appear.
This concludes the construction of the microscopic model, which is
characterized by a Hamiltonian with 26 free parameters.


\section{Fit to DFT results}

We employ the tight-binding model presented in the supplemental
material of~\citeS{S:Veenstra2014} to determine the free parameters.
The tight-binding Hamiltonian is derived from an LDA band structure
that is down-folded onto the O-$2p$ and the Ru-$4d$ orbitals and
therefore has a total of 17 bands.  The hopping integrals are
truncated at 10\,meV.  We henceforth refer to the LDA-derived
tight-binding Hamiltonian as the ``DFT model''.  For the calculation
of the linearized gap equation, the DFT model is much too large and
most of the bands are irrelevant for superconductivity.  The states at
the Fermi surface are determined by the $t_{2g}$ manifold of the
Ru-$4d$ orbitals ($d_{yz}$, $d_{xz}$, $d_{xy}$) and we fit
Eq.~\eqref{S:Eq:H0} to several quantities extracted from the DFT model
projected into this subspace.

We extract the Fermi momenta $\tilde{\bm{k}}_F$ of the DFT model and
denote the eigenvalues by $\epsilon$ and the associated eigenvectors
by $V$.  We define the following measure
\begin{equation}
  \label{eq:measure}
  S = \!\! \sum_{n=\alpha,\beta,\gamma,\: \tilde{\bm{k}}_F} \! \Bigl[
  \bigl( \epsilon^n(\tilde{\bm{k}}_F) \bigr)^2
  + \bigl( \tilde{d}_{xy}^n(\tilde{\bm{k}}_F) - d_{xy}^n(\tilde{\bm{k}}_F) \bigr)^2
  + \bigl( \tilde{p}_{\mathrm{SOC}}^n(\tilde{\bm{k}}_F) - p_{\mathrm{SOC}}^n(\tilde{\bm{k}}_F) \bigr)^2
  + \bigl( \tilde{v}_{\parallel}^n(\tilde{\bm{k}}_F) - v_{\parallel}^n(\tilde{\bm{k}}_F) \bigr)^2
  \Bigr] ,
\end{equation}
where the sum is over momenta $\tilde{\bm{k}}_F$ on the DFT Fermi
surfaces formed by the bands $n=\alpha,\beta,\gamma$,
$\epsilon^n(\bm{k})$ are the band energies,
$d_{xy}^n(\bm{k})$ is the $d_{xy}$-orbtial content,
$p_{\mathrm{SOC}}^n(\bm{k})$ is the spin polarization, and
$v_{\parallel}^n(\bm{k})$ the in-plane velocity.  Quantities with a
tilde are from the DFT model.  The $d_{xy}$-orbtial content is
determined by the corresponding eigenvector components
\begin{equation}
  d_{xy}^n(\bm{k}) = \frac{1}{2} \bigl(
  \abs{V^{n,\up}_{d_{xy},\up}(\bm{k})}^2
  + \abs{V^{n,\up}_{d_{xy},\dn}(\bm{k})}^2
  + \abs{V^{n,\dn}_{d_{xy},\up}(\bm{k})}^2
  + \abs{V^{n,\dn}_{d_{xy},\dn}(\bm{k})}^2
  \bigr) .
\end{equation}
The spin polarization is determined from the expectation value of the
atomic spin-orbit coupling Hamiltonian
$H_{\mathrm{SOC}} = \lambda_5\sigma_2 - \lambda_6\sigma_1 -
\lambda_4\sigma_3$:
\begin{equation}
  p_{\mathrm{SOC}}^n(\bm{k}) = 1 +
  \biggl[\frac{1}{2} \real\bigl(
      V^{n,\up T}(\bm{k}) H_{\mathrm{SOC}} V^{n,\up}(\bm{k})
      + V^{n,\dn T}(\bm{k}) H_{\mathrm{SOC}} V^{n,\dn}(\bm{k})
    \bigr)
  \biggr]^{1/3} .
\end{equation}
For the in-plane Fermi velocity we use a simple two-point central
finite differences stencil where $\varepsilon_{x,y}$ are small
\begin{equation}
  v_{\parallel}^n(\bm{k}) =
  \sqrt{
    \biggl\vert
      \frac%
        {\epsilon^n(\bm{k} - \varepsilon_x) - \epsilon^n(\bm{k} + \varepsilon_x)}%
        {2 \varepsilon_x}
    \biggr\vert^2
    +
    \biggl\vert
      \frac%
        {\epsilon^n(\bm{k} - \varepsilon_y) - \epsilon^n(\bm{k} + \varepsilon_y)}%
        {2 \varepsilon_y}
    \biggr\vert^2
  } .
\end{equation}
We minimize the measure \eqref{eq:measure} using the derivative-free
optimization algorithm BOBYQA of dlib~\citeS{S:dlib}.

\begin{figure}[tb]
  \centering
  \includegraphics[scale=1.2]{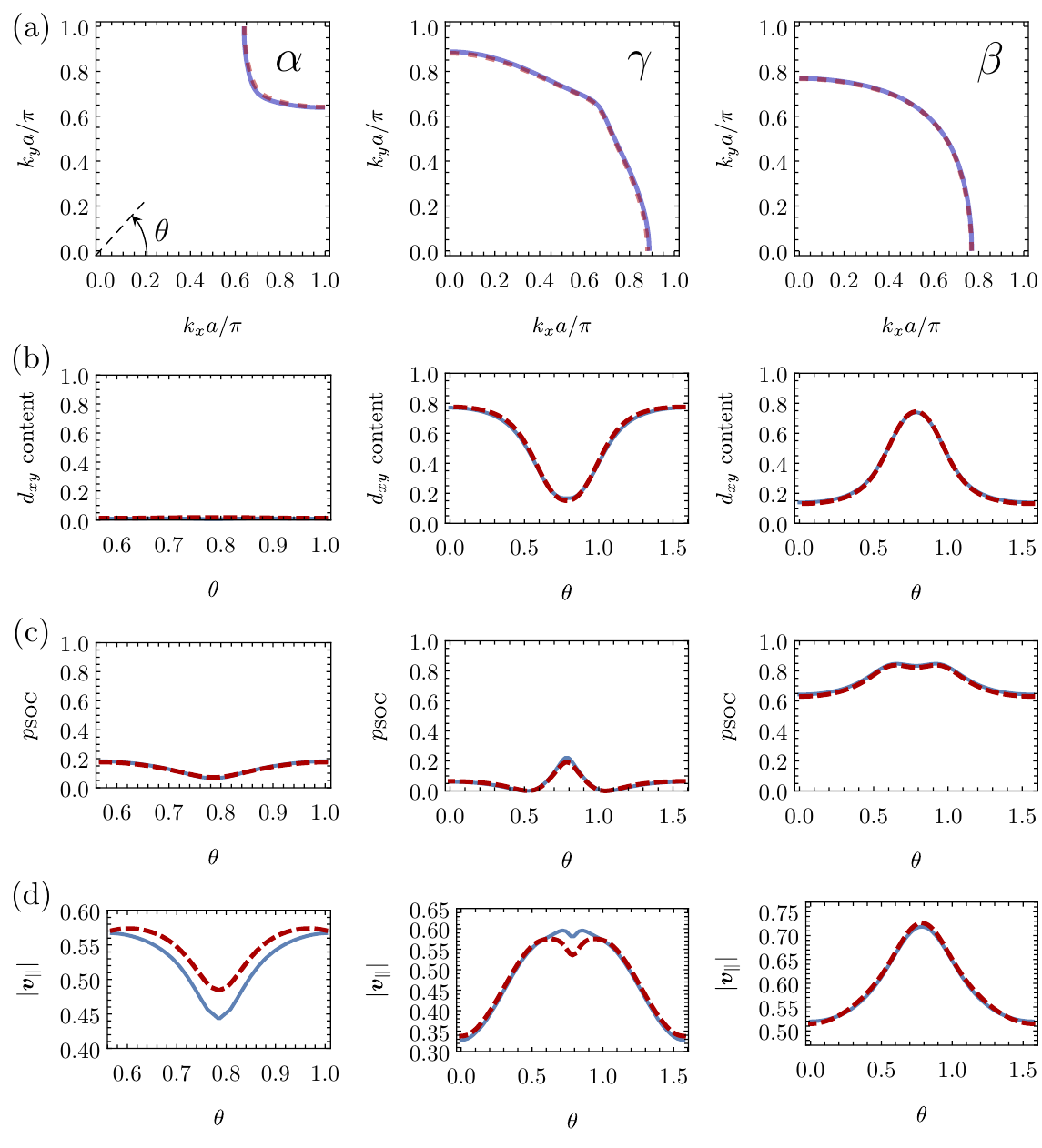}
  \caption{Comparison of the DFT model (red dashed lines) with our fit
    (blue solid lines) in the $k_z = 0$ plane. (a) Fermi surface in
    the first quadrant of the Brillouin zone. (b) $d_{xy}$-orbital
    content, (c) spin polarization, and (d) in-plane velocity as
    functions of the angle $\theta = \arctan(k_y/k_x)$ in the first
    quadrant.  The three columns pertain to the $\alpha$, $\gamma$,
    and $\beta$ band.}
  \label{fig:fit_kz0}
\end{figure}

\begin{figure}
  \centering
  \includegraphics[width=\linewidth]{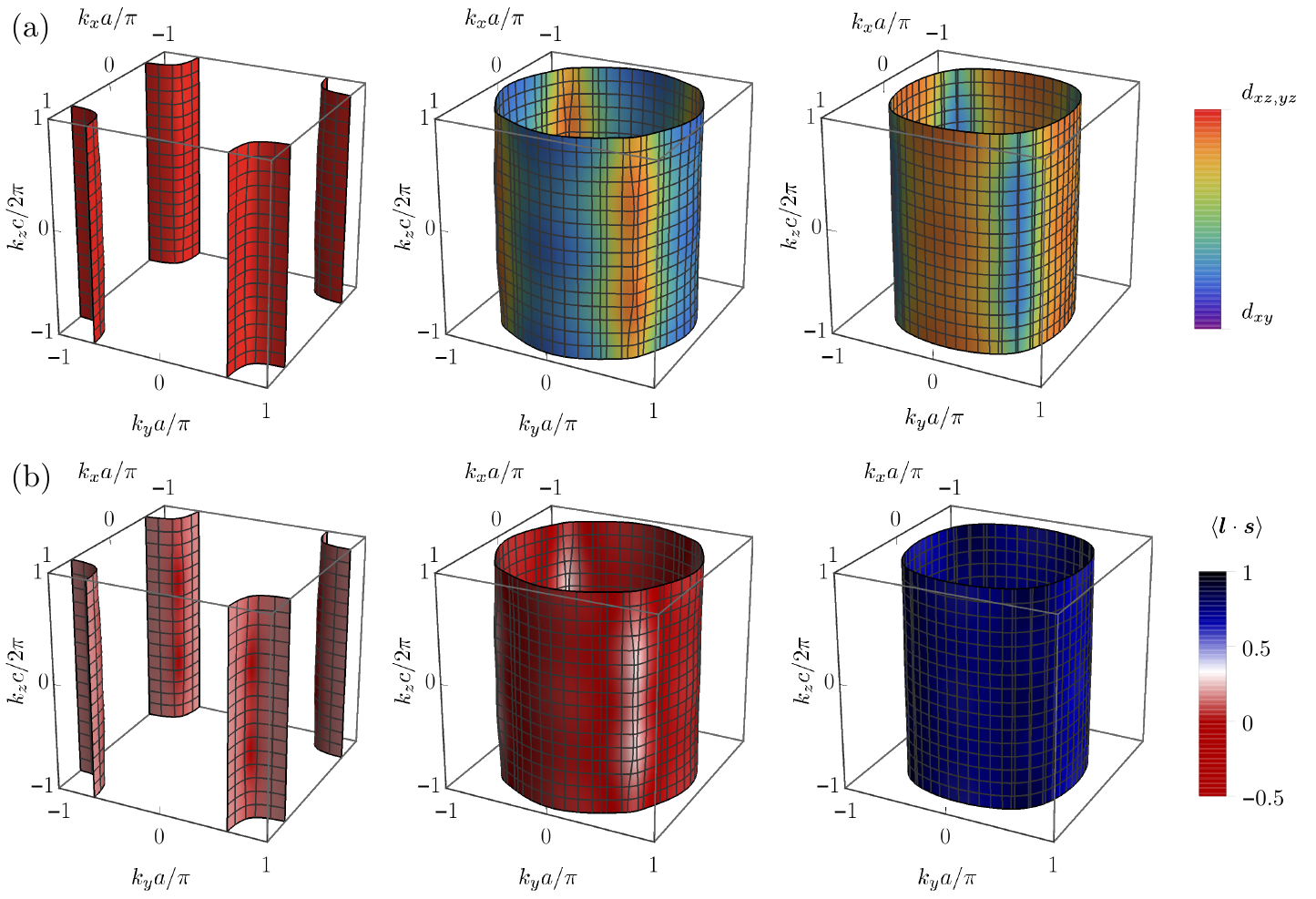}
  \caption{Full 3D Fermi surface obtained from the best fit to the DFT
    results of Ref.~\protect\citeS{S:Veenstra2014}, with color indicating (a)
    the $d_{xy}$-orbital content and (b) the spin polarization.  The
    three columns pertain to the $\alpha$, $\gamma$, and $\beta$
    band.}
  \label{fig:3d-fit}
\end{figure}

The fit yields very good agreement with the DFT model close to the
Fermi energy, including good reproduction of the $d_{xy}$-orbital
content and the spin polarization. In Fig.~\ref{fig:fit_kz0}, we
compare the result of our fit with the DFT model in the $k_z=0$ plane.
In Fig.~\ref{fig:3d-fit}, we show the full 3D Fermi surface produced
by our fit, together with the $d_{xy}$-orbital content and the spin
polarization.  The corresponding fit parameters are listed in
Tab.~\ref{tab:fit}.

\begin{table}
  \centering
   $\begin{array}{@{}*{5}{cD{c}{c}{6.8}}@{}}
     \hline
     t^{11}_{x}              & -362.4 & t^{11}_{y}   & -134    & t^{33}_{x}             & -262.4 & t^{11}_{xy}            & -44.01 & t^{11}_{xx}             & -1.021 \\
     t^{11}_{yy}             & -5.727 & t^{33}_{xy}  & -43.73  & t^{33}_{xx}            & 34.23  & t^{12}_{xy}            & 16.25  & t^{11}_{xxy}            & -13.93 \\
     t^{11}_{xyy}            & -7.52  & t^{33}_{xxy} & 8.069   & t^{12}_{xxy}           & 3.94   & \eta                   & 57.39  & \mu                     & 438.5  \\
     \mu_1                   & 218.6  & t^{11}_{z}   & -0.0228 & t^{33}_{z}             & 1.811  & t^{12}_{z}             & 19.95  & t^{13}_{z}              & 8.304  \\
     t^{11}_{zz}             & 2.522  & t^{33}_{zz}  & -3.159  & t^{\mathrm{SOC}}_{56z} & -1.247 & t^{\mathrm{SOC}}_{12z} & -3.576 & t^{\mathrm{SOC}}_{5162} & -1.008 \\
     t^{\mathrm{SOC}}_{5261} & 0.3779                                                                                                                                 \\
     \hline
   \end{array}$
   \caption{Parameters of the Hamiltonian~\eqref{S:Eq:H0} determined
     from the fit to the DFT model.  All values are in meV.}
  \label{tab:fit}
\end{table}

It is important to note that because the different sheets of the Fermi
surface have varying orbital and spin content, it is not possible to
isolate one dominant band for superconductivity. The pairing state
will in general have contributions from all three sheets.


\vspace*{3ex}

\section{Linearized gap equation}

 In this section, we outline our solution of the linearized BCS gap equation.
For convenience, we repeat the second-order expansion of the free energy as given in Eq.\ (3) of the main text,
\begin{equation}
\mathcal{F} = \frac{1}{2}\sum_{i}
\frac{1}{g_i}\, \textrm{Tr}\, [\hat{\Delta}_i^\dag \hat{\Delta}_i]
- \frac{k_B T}{2} \sum_{\bk,\omega,i,j}
\textrm{Tr}\, \big[\hat{\Delta}_i \hat{\underline{G}} \hat{\Delta}^\dag_j \hat{G}\big],
\end{equation}
where the gap functions are $\hat{\Delta}_i=\Delta_i\, \lambda_{a_i} \otimes \sigma_{b_i}\, (i\sigma_2)$ and the indices $a_i$, $b_i$, and interaction energies $g_i$ are given in Table 1 of the main text. We introduce an interaction scaling parameter $s$, and for concreteness choose the interaction energies to be given by $U=5/s$, $U'=1/s$, and $J=2/s$. Since we are interested in the weak-coupling limit we will later assume $s$ to be large.
The Green's functions and gap function are expressed in the energy eigenbasis by
\begin{align}
\hat{G} &\rightarrow U^\dagger \hat{G} U = \mathrm{diag}\left(\frac{1}{i\omega_n-\epsilon_a}\right) , \\
\hat{\underline{G}} &\rightarrow \underline{U}^\dagger \hat{\underline{G}}\, \underline{U} = \mathrm{diag}\left(\frac{1}{i\omega_n+\epsilon_a}\right) , \\
\hat{\Delta}_i &\rightarrow U^\dagger \hat{\Delta}_i\, \underline{U} ,
\end{align}
where $U$ is a unitary matrix that diagonalizes the normal-state Hamiltonian $\hat{H}_0$, $\underline{U} = (i\sigma_2)^\dagger U$, and $\epsilon_a$ are band energies. We define new gap matrices by
\begin{equation}
\Lambda_i=U^\dagger \left(\lambda_{a_i} \otimes \sigma_{b_i}\right) U .
\end{equation}
The frequency summation yields 
\begin{equation}\label{eq:matsubaraSum}
S_{ab}(\bk,\beta) = -\frac{1}{\beta} \sum_{\omega_n} \frac{1}{i\omega_n-\epsilon_a}\,
\frac{1}{i\omega_n+\epsilon_b}
= \frac{1}{2}\,\frac{\tanh(\beta\epsilon_a/2) + \tanh(\beta\epsilon_b/2)}
{\epsilon_a+\epsilon_b} ,
\end{equation}
where $\beta=1/k_BT$. 
The linearized gap equations are obtained by differentiating the free energy with respect to the gap amplitudes, $\partial \mathcal{F}/\partial\Delta_i^\ast = 0$, written explicitly as
\begin{equation}\label{eq:gapEq1}
\sum_{j} \bigg( s\frac{\delta_{ij}}{\tilde{g}_i} \textrm{Tr}\, [\Lambda_i^\dag \Lambda_i]
+\sum_{\bk,a,b}[\Lambda_i]^\ast_{ab}[\Lambda_j]_{ab}\, S_{ab} \bigg) \Delta_j
= 0 ,
\end{equation}
where $i$ and $j$ run over all gap-structure indices of a given irrep, $a$ and $b$ run over band indices, $[\Lambda_i]_{ab}$ is a matrix element of $\Lambda_i$, and $\tilde{g}_i$ is the value of $g_i$ when $s=1$.

First, consider $\epsilon_a = \epsilon_b$ intraband terms in Eq.\ (\ref{eq:gapEq1}), the $\bk$-integration is written as
\begin{equation}\label{eq:gapeq_intra}
\sum_{\bk,{ \epsilon_a=\epsilon_b}}[\Lambda_i]^\ast_{ab}[\Lambda_j]_{ab}\, S_{ab}={ \sum_{\epsilon_a=\epsilon_b}}\frac{1}{2}\int \mathrm{d}\epsilon\,F_{abij}(\epsilon)\frac{\tanh(\beta\epsilon/2)}{\epsilon},
\end{equation}
with
\begin{equation}\label{eq:Fabij}
F_{abij}(\epsilon)=\frac{1}{(2\pi)^3}{ \iint_{\epsilon_a(\bk)=\epsilon}}\mathrm{d}^2\bk\,\frac{\delta_{\epsilon_a\epsilon_b}}{{ \lVert\bm{\nabla}\epsilon_a\rVert}}[\Lambda_i]^\ast_{ab}[\Lambda_j]_{ab} .
\end{equation}
{ Use the Taylor expansion of \(F_{abij}(\epsilon)\) around the Fermi level, \(F_{abij}(0)+\epsilon F'_{abij}(0)+\dots\), the derivative of Eq.\ (\ref{eq:gapeq_intra}) with respect to $\beta$ gives}
\begin{equation}
\frac{\partial}{\partial\beta}\sum_{\bk,{ \epsilon_a=\epsilon_b}}[\Lambda_i]^\ast_{ab}[\Lambda_j]_{ab}\, S_{ab}
=\frac{1}{\beta}{ \sum_{\epsilon_a=\epsilon_b}} F_{abij}(0)+O\left(\frac{1}{\beta^3}\right) .
\end{equation}
Note that when this is integrated with respect to $\beta$ it yields the  $\log\beta$ divergence in Eq.\ (\ref{eq:gapeq_intra}). Next, consider the $\epsilon_a\neq\epsilon_b$ interband terms, as $\beta\to\infty$, $S_{ab}$ converges to $\theta(\epsilon_a\epsilon_b)/|\epsilon_a+\epsilon_b|$ , where \(\theta\) is the Heaviside step function. Because this is a bounded function, there is no divergence in the interband contributions.
In Eq.\ (\ref{eq:gapEq1}) a non-trivial solution for the gap amplitudes $\Delta_j$ is found by considering $i$ and $j$ as matrix indices and taking the corresponding $6\times 6$ matrix to be  singular. Including both the  intraband and interband contributions, the critical \(\beta_c\) satisfies
\begin{equation}\label{eq:gapEq2}
\det\bigg[s\frac{\delta_{ij}}{\tilde{g}_i} \textrm{Tr}\, [\Lambda_i^\dag \Lambda_i] + \log\beta_c \sum_{\epsilon_a=\epsilon_b} F_{abij}(0) +C_{ij}(\beta_c)\bigg] = 0 ,
\end{equation}
where  \(C_{ij}(\beta)\) is the portion of \(\sum_{\bk,a,b}[\Lambda_i]^\ast_{ab}[\Lambda_j]_{ab}\, S_{ab}\) that remains after removing the $\log\beta$ divergent term. 
By definition, $C_{ij}(\beta)$ is convergent as $\beta\to\infty$, so the last term in the determinant can be ignored when $s$ is sufficiently large. More explicitly, in the weak-coupling limit \(s\rightarrow\infty\), $T_c$ is given in the form
\begin{equation}\label{eq:Tcrelation}
\log T_c(s) \approx -ms+{ \delta},
\end{equation}
where { $\delta$ is a constant} and $m$ is the smallest  $\log \beta_c/s$ solution found when $C_{ij}=0$.

Different channels (irreps) have different values of $m$, and the channel with the smallest $m$ is the leading instability in the weak-coupling limit. Note that the definition of $m$ does not depend on $C_{ij}$ and all the interband contributions go into $C_{ij}$. Thus we can drop the interband terms in Eq.\ (\ref{eq:gapEq1}). This changes \(\delta\) but does not change  $m$. The resultant expression is
\begin{equation}\label{eq:TcNum}
\det\bigg[s\frac{\delta_{ij}}{\tilde{g}_i} \textrm{Tr}\, [\Lambda_i^\dag \Lambda_i]
+ \sum_{\bk,\epsilon_a=\epsilon_b}[\Lambda_i]^\ast_{ab}[\Lambda_j]_{ab}\, S_{ab}\bigg] = 0 ,
\end{equation}
which  is the equation we solve numerically. The $\log\beta$ divergence originates from momenta near the Fermi surface, so we carry out the $\bk$-integration on adaptive meshes with finer resolution near the Fermi surface.
We obtain $\log\beta_c$ for several values of $s$ and use linear regression to get the slope, which  determines $m$. If the values of $\log\beta_c$ at the sampling points are not linear in $s$ we sample larger $s$ values until we encounter linear behavior. In our calculation, an equidistant set of four sampling points is used for a linear regression and their $R^2$ measures are always greater than $0.999$. Using this procedure, we get the slope $m$ for each pairing channel and determine the leading instability at each point in the phase diagram displayed in Fig.\ 1(a) in the main text. While this procedure may seem more elaborate than a direct solution of Eq.\ (\ref{eq:gapEq2}) with $C_{ij}=0$,
it allows us to verify Eq.\ (\ref{eq:Tcrelation}) showing that our solution is in the weak coupling limit.


\section{Superconducting-fitness analysis}

In this section, we present details of the superconducting-fitness analysis. We start with the more realistic three-orbital model and then consider an effective two-orbital model, which dramatically simplifies the analysis but gives consistent results.


\subsection{Complete 3D three-orbital model}

In previous works \citeS{S:RamiresPRB2016, S:RamiresPRB2018}, a proof of the direct relation between the superconducting-fitness measures $\hat{F}_C(\bk)$ and $\hat{F}_A(\bk)$ (defined below) and the superconducting critical temperature was provided for the one- and two-orbital scenario. The first measure,
\begin{equation} \label{Eq:FCDef}
\hat{F}_C(\bk) =  \tilde{H}_0(\bk) \tilde{\Delta}(\bk)
  - \tilde{\Delta}(\bk)  \tilde{H}^*_0(-\bk),
\end{equation}
quantifies how incompatible a given gap structure is for a specific normal state, namely, how much inter-band pairing there is. Here, $\tilde{H}_0(\bk) = \hat{H}_0(\bk) - h_{00}(\bk)\, \lambda_0\otimes \sigma_0$ and we have defined a normalized gap matrix $\tilde{\Delta}(\bk) = \hat{\Delta}(\bk)/|\hat{\Delta}(\bk)|$ such that average over the normal-state Fermi surface is $\langle \tilde{\Delta}(\bk)\tilde{\Delta}^\dagger(\bk) \rangle_\text{FS} = \mathbb{1}$. The second measure,
\begin{equation} \label{Eq:FADef}
\hat{F}_A(\bk) =  \tilde{H}_0(\bk) \tilde{\Delta}(\bk)
  + \tilde{\Delta}(\bk)  \tilde{H}^*_0(-\bk),
\end{equation}
quantifies how much intra-band pairing there is, or what fraction of the gap survives upon projection onto the Fermi surface. For the two-orbital scenario, these measures  satisfy $\langle\text{Tr}\, \hat F_A^\dagger(\bk) \hat F_A(\bk) + \text{Tr}\, \hat F_C^\dagger(\bk) \hat F_C(\bk)\rangle_{\text{FS}} = 1$, up to normalization of the normal-state Hamiltonian, which highlights their complementarity.
The proof of this relation relies on the fact that the matrices associated with the orbital DOF are Pauli matrices for the two-orbital scenario  and therefore form a totally anticommuting set, which greatly simplifies the calculations. On the other hand, for $n>2$ orbitals, the basis matrices are the generators of $\mathrm{SU}(n)$, which do not form a totally anticommuting set and therefore do not allow a direct generalization of this relation for models with more than two orbitals. However, the physical meaning of $\hat{F}_C(\bk)$ and $\hat{F}_A(\bk)$ is preserved within some approximations, as discussed below.

For the three-orbital situation, the corresponding superconducting-fitness functions can be identified as
\begin{equation}
 \hat{F}^{3orb}_{A,C}(\bk) = [\hat{H}_0(\bk)]^2 \tilde{\Delta}(\bk) \pm \tilde{\Delta}(\bk)[\hat{H}^*_0(-\bk)]^2.
\end{equation}
Given that $[A^2, B]_{\pm} = A[A,B]\mp [A,B]A$, the core of the analysis still depends on the original form of the superconducting-fitness functions. Therefore, we use the form linear in $\hat{H}_0(\bk)$ to get some insight. Below, we will see that a simplified  two-orbital model, for which the linear version of the fitness functions is valid rigorously, corroborates our analysis.
We summarize the results for the complete three-orbital problem in Tables \ref{Tab:FA} and \ref{Tab:FC}. The first row gives the irrep of each term in the normal-state Hamiltonian displayed in the second  row as $h_{ab}$. The first column gives the irrep of each order parameter displayed in the second column following the notation $[a,b]$ corresponding to $\hat{\Delta} = \lambda_a \otimes \sigma_b\, (i\sigma_2)$. The third column gives the interaction stemming from the Hubbard-Kanamori Hamiltonian for each channel. Finally, the numerical entries in the tables correspond to $\text{Tr}\, \hat F_{A,C}^\dagger \hat F_{A,C} = \sum_{cd} (\text{table entry})\, |h_{cd}|^2$.

\begin{table}[tb]
\begin{center}
\def\arraystretch{1.25}
\begin{tabular}{|c|c|c|c|c|c|c|c|c|c|c||c|c||c|c||c|c||}
\cline{4-17}
 \multicolumn{3}{c|}{}& \multicolumn{3}{c|}{$A_{1g}$} & $A_{2g}$ & \multicolumn{2}{c|}{$B_{1g}$} & \multicolumn{2}{c||}{$B_{2g}$} & \multicolumn{6}{c||}{$E_{g}$}\\
\hline
Irrep & $[a,b]$ & Interac.
& $\boldsymbol{h_{80}}$ & $\boldsymbol{h_{43}}$ &  $\boldsymbol{h_{52-61}}$& $h_{51+62}$ & $\boldsymbol{h_{70}}$ & $h_{52+61}$ & $\boldsymbol{h_{10}}$ & $h_{51-62}$ & $h_{20}$ & $h_{30}$ & $h_{41}$ & $h_{42}$ & $h_{53}$ & $h_{63}$ \\
\hline
\multirow{4}{*}{$A_{1g}$} & $[0,0]$ & $U+2J^\prime$ &$ \frac{32}{3}$ & $\frac{32}{3}$ &$ \frac{64}{3}$ & $\frac{128}{3}$ & $\frac{32}{3}$ & $\frac{128}{3}$ &$ \frac{32}{3} $&$ \frac{64}{3}$ &$ \frac{32}{3} $ &$ \frac{32}{3}$ &$ \frac{32}{3}$ &$ \frac{32}{3}$ & $\frac{32}{3}$ & $\frac{32}{3}$\\\cline{2-17}
 & $[8,0]$ & $U-J^\prime$ & 16 &$ \frac{16}{3} $& $\frac{8}{3}$ & $\frac{16}{3}$ &$ \frac{16}{3}$ & $\frac{16}{3}$ & $\frac{16}{3} $&$ \frac{8}{3}$ & $ \frac{4}{3} $& $\frac{4}{3}$ &$ \frac{16}{3} $& $\frac{16}{3}$ &$ \frac{4}{3}$ &$ \frac{4}{3}$ \\\cline{2-17}
 & $[4,3]$ & $\boldsymbol{U^\prime-J}$ & $\frac{16}{3}$ & 16 & 8 & 16 & 0 & 16 & 0 & 8 & 4 & 4 & 0 & 0 & 4 & 4 \\\cline{2-17}
 & $[5,2]-[6,1]$ & $\boldsymbol{U^\prime-J}$ & $\frac{5}{3}$ & 3 & 24 & 8 & 3 & 8 & 3 & 8 & 2 & 2 & 4 & 4 & 2 & 2\\\hline
 $A_{2g}$ & $[5,1]+[6,2]$ & $\boldsymbol{U^\prime-J}$ & $\frac{5}{3}$ & 3 & 0 & 32 & 3 & 32 & 3 & 8 & 2 & 2 & 4 & 4 & 2 & 2   \\\hline
\multirow{2}{*}{ $B_{1g}$} & $[7,0]$ & $U-J^\prime$ & $\frac{16}{3}$ & 0 & 8 & 16 & 16 & 16 & 0 & 8 & 4 & 4 & 0 & 0 & 4 & 4 \\\cline{2-17}
 & $[5,2]+[6,1]$ & $\boldsymbol{U^\prime-J}$ & $\frac{5}{3} $& 3 & 8 & 32 & 3 & 32& 3 & 0 &2 & 2 & 4 & 4 & 2 & 2  \\\hline
 \multirow{2}{*}{$B_{2g} $} & $[1,0]$ & $U^\prime+J$ & $\frac{16}{3}$ & 0 & 8 & 16 & 0 & 16 & 16 & 8 & 4 & 4 & 0 & 0 & 4 & 4  \\\cline{2-17}
 & $[5,1]-[6,2]$ & $\boldsymbol{U^\prime-J}$ & $\frac{5}{3}$ & 3 & 8 & 8 & 3 & 8 & 3 &24 & 2 & 2 & 4 & 4 & 2 & 2 \\\hline
 \multirow{2}{*}{$E_g$} & $+[3,0]$ & \multirow{2}{*}{$U^\prime+J$ }& $\frac{4}{3} $& 4 & 4 & 8 & 4 & 8 & 4 & 4 & 4 & 16 & 4 & 4 & 4 & 0\\\cline{2-2}\cline{4-17}
   & $-[2,0]$ & & $ \frac{4}{3}$ & 4 & 4 & 8 & 4 & 8 & 4 & 4 & 16 & 4 & 4 & 4 & 0 & 4 \\\hline
  \multirow{2}{*}{$E_g$}  & $+[4,2]$ &\multirow{2}{*}{$\boldsymbol{U^\prime-J}$} & $\frac{16}{3}$ & 0 & 8 & 16 & 0 & 16 & 0 & 8 & 4 & 4 & 0 & 16 & 4 & 4 \\\cline{2-2}\cline{4-17}
   & $-[4,1]$ & & $\frac{16}{3}$ & 0 & 8 & 16 & 0 & 16 & 0 & 8 & 4 & 4 & 16 & 0 & 4 & 4 \\\hline
 \multirow{2}{*}{$E_g$}  & $+[5,3]$ & \multirow{2}{*}{$\boldsymbol{U^\prime-J}$}&$ \frac{4}{3} $& 4 & 4 & 8 & 4 & 8 & 4 & 4 & 0 & 4 & 4 & 4 & 16 & 4 \\\cline{2-2}\cline{4-17}
  & $+[6,3]$ &  & $\frac{4}{3}$ & 4 & 4 & 8 & 4 & 8 & 4 & 4 & 4 & 0 & 4 & 4 & 4 & 16 \\\hline
\end{tabular}
\end{center}
\caption{Superconducting-fitness measure $\hat F_A$ for the 3D
  three-orbital model for Sr$_2$RuO$_4$. The first column  gives
  the irreps of the order parameters associated with matrix form
  $[a,b]$ (second column) and the third column displays the local
  interaction in the respective channel, where the potentially
    attractive channels are highlighted in boldface. Columns 4--17
  give the results for the fitness function such that $\text{Tr}\,
   \hat F_A^\dagger \hat F_A = \sum_{cd} (\text{table entry})\,
  |h_{cd}|^2$, for each term $h_{cd}$ in the normal-state Hamiltonian,
   indicated in the second row with the associated irrep given in
    the first row. We highlight in boldface the $h_{cd}$ terms which
  are present in the standard 2D models for Sr$_2$RuO$_4$, while the
  terms in normal typeface are either momentum-dependent SOC or
    interlayer couplings.}
\label{Tab:FA}
\end{table}

\begin{table}[tb]
\begin{center}
\def\arraystretch{1.25}
\begin{tabular}{|c|c|c|c|c|c|c|c|c|c|c||c|c||c|c||c|c||}
\cline{4-17}
 \multicolumn{3}{c|}{}& \multicolumn{3}{c|}{$A_{1g}$} & $A_{2g}$ & \multicolumn{2}{c|}{$B_{1g}$} & \multicolumn{2}{c||}{$B_{2g}$} & \multicolumn{6}{c||}{$E_{g}$}\\
\hline
Irrep & $[a,b]$ & Interac.
& $\boldsymbol{h_{80}}$ & $\boldsymbol{h_{43}}$ & $\boldsymbol{h_{52-61}}$ & $h_{51+62}$ & $\boldsymbol{h_{70}}$  & $h_{52+61}$ & $\boldsymbol{h_{10}}$ & $h_{51-62}$ & $h_{20}$ & $h_{30}$ & $h_{41}$ & $h_{42}$ & $h_{53}$ & $h_{63}$ \\
\hline
\multirow{4}{*}{$A_{1g}$} & $[0,0]$ & $U+2J^\prime$ & 0 & 0 & 0 & 0 & 0 & 0 & 0 & 0 & 0 & 0 & 0 & 0 & 0 & 0 \\\cline{2-17}
 & $[8,0]$ & $U-J^\prime$ & 0 & 0 & 24 & 48 & 0 & 48 & 0 & 24 & 12 & 12 & 0 & 0 & 12 & 12 \\\cline{2-17}
 & $[4,3]$ & $\boldsymbol{U^\prime-J}$& 0 & 0 & 8 & 16 & 16 & 16 & 16 & 8 & 4 & 4 & 16 & 16 & 4 & 4 \\\cline{2-17}
& $[5,2]-[6,1]$ & $\boldsymbol{U^\prime-J}$  & 11 & 3 & 0 & 32 & 3 & 32 & 3 & 8 & 8 & 8 & 4 & 4 & 8 & 8  \\\hline
 $A_{2g}$ & $[5,1]+[6,2]$ & $\boldsymbol{U^\prime-J}$ & 11 & 3 & 24 & 8 & 3 & 8 & 3 & 8 & 8 & 8 & 4 & 4 & 8 & 8   \\\hline
\multirow{2}{*}{ $B_{1g}$ } & $[7,0]$ & $U-J^\prime$ & 0 & 16 & 8 & 16 & 0 & 16 & 16 & 8 & 4 & 4 & 16 & 16 & 4 & 4  \\\cline{2-17}
& $[5,2]+[6,1]$ & $\boldsymbol{U^\prime-J}$& 11 & 3 & 8 & 8 & 3 & 8 & 3 & 24 & 8 &8& 4 & 4 & 8 & 8 \\\hline
 \multirow{2}{*}{$B_{2g}$} & $[1,0]$ & $U^\prime+J$ & 0 & 16 & 8 & 16 & 16 & 16 & 0 & 8 & 4 & 4 & 16 & 16 & 4 & 4  \\\cline{2-17}
& $[5,1]-[6,2]$ & $\boldsymbol{U^\prime-J}$  & 11 & 3 & 8 & 32 & 3 & 32 & 3 & 0 & 8 & 8 & 4 & 4 &8 & 8 \\\hline
  \multirow{2}{*}{$E_g$} & $+[3,0]$ & \multirow{2}{*}{$U^\prime+J$ } & 12 & 4 & 20 & 40 & 4 & 40 & 4 & 20 & 4 & 0 & 4 & 4 & 4 & 16 \\\cline{2-2}\cline{4-17}
  & $-[2,0]$ & & 12 & 4 & 20 & 40 & 4 & 40 & 4 & 20 & 0 & 4 & 4 & 4 & 16 & 4\\\hline
  \multirow{2}{*}{$E_g$} & $+[4,2]$ &\multirow{2}{*}{$\boldsymbol{U^\prime-J}$} & 0 & 16 & 8 & 16 & 16 & 16 & 16 & 8 & 4 & 4 & 16 & 0 & 4 & 4  \\\cline{2-2}\cline{4-17}
  & $-[4,1]$ & & 0 & 16 & 8 & 16 & 16 & 16 & 16 & 8 & 4 & 4 & 0 & 16 & 4 & 4  \\\hline
 \multirow{2}{*}{$E_g$} & $+[5,3]$ &\multirow{2}{*}{$\boldsymbol{U^\prime-J}$} & 12 & 4 & 20 & 40 & 4 & 40 & 4 & 20 & 16 & 4 & 4 & 4 & 0 & 4 \\\cline{2-2}\cline{4-17}
  & $+[6,3]$ & & 12 & 4 & 20 & 40 & 4 & 40 & 4 & 20 & 4 & 16 & 4 & 4 & 4 & 0  \\\hline
\end{tabular}
\end{center}
\caption{Superconducting-fitness measure  $\hat F_C$ for the 3D three-orbital model for Sr$_2$RuO$_4$. The same notation as in Table \ref{Tab:FA} has been used.}
\label{Tab:FC}
\end{table}

Note that the order parameters with a potentially attractive interaction $U'-J$ are $[4,3]$ and $[5,2]-[6,1]$ in $A_{1g}$, $[5,1]+[6,2]$ in $A_{2g}$, $[5,2]+[6,1]$ in $B_{1g}$, $[5,1]-[6,2]$ in $B_{2g}$, and finally  $\{[4,2],-[4,1]\}$ and $\{[5,3],[6,3]\}$ in $E_g$. All these order parameters are associated with spin-triplet states. If we focus first on the largest terms in the normal-state Hamiltonian, namely $h_{80}$ and $h_{70}$ (intra-orbital hopping), $h_{10}$ (inter-orbital hopping), and $h_{43}$ and $h_{52-61}$ (atomic SOC), we conclude from Tables \ref{Tab:FA} and \ref{Tab:FC} that, among the one-dimensional irreps, the most stable state should be in the $A_{1g}$ channel since these states are associated with larger entries for $\hat F_A$ and smaller entries for $\hat F_C$.

Considering now the two-dimensional order parameters, for 
  $\{[4,2],-[4,1]\}$, we find that the terms stabilizing it, i.e., the ones with the largest contribution to
 $\hat F_A$, are $h_{51+62}$, $h_{52+61}$, and 
  $\{h_{42},-h_{41}\}$,  all associated with momentum-dependent SOC. However, these terms contribute with  the same value to the detrimental fitness measure $\hat F_C$, suggesting that they overall do not favor this pairing state. On the other hand, the two-dimensional order parameter $\{[5,3],[6,3]\}$ is stabilized by $h_{51+62}$, $h_{52+61}$, and $\{h_{53},h_{63}\}$. Again, the  terms $h_{51+62}$ and $h_{52+61}$ contribute with a  large value to $\hat F_C$. On the other hand, the terms $\{h_{53},h_{63}\}$ contribute only moderately. This analysis suggests that the $\{[5,3],[6,3]\}$ channel should be the one driving the superconducting instability in the $E_g$ channel and can be stabilized by large terms $\{h_{53},h_{63}\}$.

From this analysis, we can understand the tendencies observed in the numerical results as follows: the order parameters in $A_{1g}$, in particular $[5,2]-[6,1]$, are strongly stabilized by atomic SOC, in particular by the term $h_{52-61}$ in the normal-state Hamiltonian, such that reducing the magnitude of  this coupling is expected to weaken the superconducting instability in this channel. Moreover, the terms $\{h_{53},h_{63}\}$ primarily suppress the order parameter in this channel since  their contribution to $\hat F_C$ is larger than the one to $\hat F_A$. In contrast, the $E_g$ order parameters, in particular for $\{[5,3],[6,3]\}$,  are primarily stabilized by $\{h_{53},h_{63}\}$ since for these  terms the contribution to $\hat F_A$ is larger than the one to $\hat F_C$, while atomic SOC is clearly detrimental. This analysis suggests that by reducing the atomic SOC and enhancing the terms $\{h_{53},h_{63}\}$ associated with nonlocal SOC even in momentum, the ground state should change from $A_{1g}$ to $E_g$.


\subsection{Effective two-orbital model in the $k_xk_z$-plane}

Sufficiently far from the Brillouin-zone diagonals $k_y=\pm
  k_x$, the bands close to the Fermi energy are dominated by only two
  of the Ru $d$-orbitals. For concreteness, here we consider the
  $k_xk_z$-plane, but our conclusions remain qualitatively valid for
  general $\bk$, except close to  $k_y=\pm k_x$.

 In the
  $k_xk_z$-plane, the dominant orbitals at the Fermi energy are $d_{xz}$ and
  $d_{xy}$. Projecting into this subspace, we obtain an
    effective two-orbital Hamiltonian which is parametrized by
\begin{eqnarray}\label{Eq:H02orb}
\hat{H}_{2\,\text{orb}}(\bk) = \tilde{h}_{ab}(\bk)\, \tau_a \otimes \sigma_b,
\end{eqnarray}
where  the $\tilde{h}_{ab}(\bk)$ are real functions of momentum,
$\tau_a$ and $\sigma_b$ are Pauli matrices for $a,b = 1,2,3$ and
  the $2\times 2$ identity matrix for $a,b=0$, encoding the orbital
  and the spin DOF, respectively.
There are, in principle, 16 parameters $\tilde{h}_{ab}(\bk)$
but in  the presence of time-reversal and inversion symmetries
these are constrained to only six, including the term
proportional to the identity. The symmetry-allowed terms are listed in
Table \ref{Tab:H02}; we classify them in terms of the irreps of
$D_{2h}$, which is the little group for $D_{4h}$
 in the $k_xk_z$-plane.
Analogously, we can parametrize the $s$-wave gap matrices in the orbital basis as
\begin{eqnarray}\label{Eq:D}
\hat{\Delta} = d_0\, \tau_a \otimes \sigma_b\, (i\sigma_2).
\end{eqnarray}
 The irreps associated with each $[a,b]$ combination are the same as for the normal-state Hamiltonian, given in the first two columns of Table~\ref{Tab:H02}.

\begin{table}[tb]
\begin{center}
\def\arraystretch{1.25}
    \begin{tabular}{| c | c | c | c | c || c | }
    \hline
     Irrep &  $(a,b)$ & Type & Basis & Value  in $k_xk_z$-plane & Three-orbital model \\ \hline
      \multirow{3}{*}{$A_{g}$} &  $(0,0)$ & intra-orbital hopping & \multirow{3}{*}{$1,x^2,y^2,z^2$} &  \multirow{3}{*}{finite } &  $(0,0)$ in $A_{1g}$ \\\cline{2-3}\cline{6-6}        
    &  $(3,0)$ & intra-orbital hopping &  &  &  $(8,0)$ in $A_{1g}$ \\ \cline{2-3} \cline{6-6}
    &  $(2,1)$ & atomic SOC & & &  $(6,1)$ in $A_{1g}/B_{1g}$ \\ \hline
      $B_{1g}$ &  $(2,2)$ & $\bk$-SOC & $xy$ & 0 & $(6,2)$ in $A_{2g}/B_{2g}$ \\ \hline
      $B_{2g}$ &  $(2,3)$ & $\bk$-SOC & $xz$ & finite &  $(6,3)$ in $E_g$ \\ \hline
      $B_{3g}$ &  $(1,0)$ & inter-orbital hopping & $yz$ & 0 & $(3,0)$ in $E_g$ \\ \hline
    \end{tabular}
\end{center}
\caption{List of the  six symmetry-allowed terms in the
  effective two-orbital normal-state Hamiltonian
  $\hat{H}_{2\,\text{orb}}(\bk)$ given by Eq.\ (\ref{Eq:H02orb}). For
  each $(a,b)$, the basis function $\tilde{h}_{ab}(\bk)$ should
  transform according to a specific irrep of $D_{2h}$ and can be associated with different physical processes (``Type''). The table also  gives associated basis functions  and provides information on whether they are finite or zero in the $k_xk_z$-plane and on the associated term in the original three-orbital model.}
\label{Tab:H02}
\end{table}

\begin{table}[tb]
\begin{center}
\def\arraystretch{1.25}
\begin{tabular}{|c|c|c||c|c|c|c|c||c|c|c|c|c||}
\cline{4-13}
 \multicolumn{3}{c||}{} & \multicolumn{5}{c||}{$\hat F_A$} & \multicolumn{5}{c||}{ $\hat F_C$} \\
\cline{4-13}
 \multicolumn{3}{c||}{}& \multicolumn{2}{c|}{$A_{g}$} &$ B_{1g}$ & $B_{2g}$ &$ B_{3g}$ & \multicolumn{2}{c|}{$A_{g}$} & $B_{1g}$ & $B_{2g}$ & $B_{3g}$ \\
\hline
Irrep & $[a,b]$ & Interac. &  $\boldsymbol{\tilde{h}_{30}}$ & $\boldsymbol{ \tilde{h}_{21}}$ & $\tilde{h}_{22}$& $\tilde{h}_{23}$& $\tilde{h}_{10}$ &  $\boldsymbol{\tilde{h}_{30}}$ & $ \boldsymbol{ \tilde{h}_{21}}$ & $\tilde{h}_{22}$& $\tilde{h}_{23}$& $\tilde{h}_{10}$ \\
\hline
\multirow{3}{*}{$A_{g}$}& $[0,0]$ & $U+2J^\prime$  & 1 & 1 & 1 & 1 & 1 &  0 & 0 & 0 & 0 & 0 \\ \cline{2-13}
 & \text{[3,0]} & $U-J^\prime$  & 1 & 0 & 0 & 0 & 0 & 0 & 1 & 1 & 1 & 1 \\ \cline{2-13}
 & \text{[2,1]} & $\boldsymbol{U^\prime-J}$ & 0 & 1 & 0 & 0 & 0 & 1 & 0 & 1 & 1 & 1 \\ \hline
$ B_{1g}$ & \text{[2,2]} & $\boldsymbol{U^\prime-J}$  & 0 & 0 & 1 & 0 & 0  & 1 & 1 & 0 & 1 & 1 \\ \hline
 $B_{2g}$ & \text{[2,3]} & $\boldsymbol{U^\prime-J}$ & 0 & 0 & 0 & 1 & 0 & 1 & 1 & 1 & 0 & 1 \\ \hline
$ B_{3g} $& \text{[1,0]} &$U^\prime+J$ & 0 & 0 & 0 & 0 & 1 & 1 & 1 & 1 & 1 & 0 \\ \hline
\end{tabular}
\end{center}
\caption{ Superconducting-fitness analysis for the effective
  two-orbital model in the  $k_xk_z$-plane. The first column
   gives  the
  irrep of $D_{2h}$ of the order parameter parametrized by the matrices $[a,b]$
  (second column), the third column displays the local interaction in
  the respective channel, where the potentially attractive
    channels are highlighted in boldface. Columns 4--8 give the
  results for the fitness function  $\hat F_A$ such that
  $\text{Tr}\,  \hat F_A^\dagger(\bk) \hat F_A(\bk) = \sum_{cd}
  (\text{table entry})\, |\tilde{h}_{cd}(\bk)|^2$, for each term 
    $[c,d]$ in the normal-state Hamiltonian. Analogously, columns
  9--13 give the results for the fitness function  $\hat
    F_C$. We highlight in boldface the $\tilde{h}_{cd}$ terms which
  are usually present in 2D  models, while the terms in
  normal typeface describe momentum-dependent SOC or interlayer hopping.}
\label{Tab:FAC2}
\end{table}

The superconducting-fitness analysis,  which is summarized in
Table \ref{Tab:FAC2}, is very much simplified in the two-orbital
scenario since  the symmetry-allowed matrices form a totally
anticommuting  set. From  the table, one can see that the
results concerning $\hat F_A(\bk)$ and $\hat F_C(\bk)$ are
complementary. Note that the trivial order parameter, $[0,0]$, is
stabilized by all the terms in the Hamiltonian while the remaining
order parameters of the form $[a,b]$ need the associated term 
  $\tilde{h}_{ab}$ in the Hamiltonian to develop a weak-coupling
instability. There is an attractive interaction in the
  orbital-singlet spin-triplet channels $[2,b]$. The order parameter
$[2,1]$ in $A_g$ is stabilized by the atomic SOC term
$\tilde{h}_{21}$. 
The other two potentially attractive channels $[2,2]$ in $B_{1g}$ and
$[2,3]$ in $B_{2g}$ are stabilized by $\tilde{h}_{22}$ and
$\tilde{h}_{23}$, respectively. Note,  however, that
$\tilde{h}_{22}$ is zero  in the $k_xk_z$-plane (also also in the
  equivalent $k_yk_z$-plane), which should significantly reduce
  the stability of this state. We are then left with $[2,1]$ in $A_g$ and $[2,3]$ in $B_{2g}$ as good candidates: For  strong atomic SOC $\tilde{h}_{21}$, the $A_{g}$ channel should be the most stable, whereas for $\tilde{h}_{23}>\tilde{h}_{21}$, the $B_{2g}$ channel becomes the most robust.

We now connect this discussion with the results of the 
  three-orbital analysis above. The order parameter $[2,1]$ 
in the two-orbital model corresponds to both
$[5,2]-[6,1]$ in $A_{1g}$ and $[5,2]+[6,1]$ in $B_{1g}$ of the
three-orbital model, whereas $[2,3]$ in the two-orbital model corresponds to
$\{[5,3],[6,3]\}$ in $E_g$.  As discussed in the main text, the
  leading  pairing
instabilities  are in the $E_g$ and $A_{1g}$ channels, whereas the
$B_{1g}$ channel is the subleading instability over much of the region
where the $A_{1g}$ channel is dominant. The fact that the $B_{1g}$
channel is a subleading instability is not surprising, since it must
go through a zero as one moves along the Fermi surface from the
$k_xk_z$- to the $k_yk_z$-plane, whereas the $A_{1g}$ channel
maintains a full gap. Since the attractive interactions in both
channels are the same, the $A_{1g}$ state will
be favored over $B_{1g}$.

The fact that atomic SOC favors the $A_{1g}$ channel, while 
  increasing the $\{h_{53},h_{63}\}$ terms can stabilize an  $E_g$
  state, is consistent with the  numerical analysis presented
in the main text. A naive interpretation of the two-orbital
  model implies that the $E_g$ state is stabilized over the $A_{1g}$
  when $\tilde{h}_{23}>\tilde{h}_{21}$. However, we numerically find in the
  full three-orbital model that the
  condition is closer to $\{h_{53},h_{63}\} \gtrsim h_{52-61}/4$. 
    This
  discrepancy  reflects the fact that the two-orbital model is not
  valid over the entire Brillouin zone.
 Nevertheless, the
  two-orbital model accurately identifies the terms which
  stabilize the $E_g$ state over the $A_{1g}$.


\bibliographyS{references}  

\end{document}